\documentclass[acmlarge]{acmart}
\AtBeginDocument{%
  }

\setcopyright{acmlicensed}
\copyrightyear{2018}
\acmYear{2018}
\acmDOI{XXXXXXX.XXXXXXX}
\acmConference[Proc. ACM Interact. Mob. Wearable Ubiquitous Technol.]{Make sure to enter the correct
  conference title from your rights confirmation email}{June 03--05,
  2018}{Woodstock, NY}
\acmISBN{978-1-4503-XXXX-X/2018/06}

\author{Weiye Xu}
\orcid{0000-0001-5031-8154}
\affiliation{%
  \institution{Tsinghua University}
  \city{Beijing}
  \country{China}
}
\email{xuwy24@mails.tsinghua.edu.cn}

\author{Zhang Jiang}
\orcid{0009-0003-0932-7907}
\affiliation{%
  \institution{Tsinghua University}
  \city{Beijing}
  \country{China}
}
\email{jiangzhang20030102@gmail.com}

\author{Siqi Zheng}
\orcid{0009-0002-6787-4223}
\affiliation{%
  \institution{Tsinghua University}
  \city{Beijing}
  \country{China}
}
\email{ymzqwq@gmail.com}

\author{Xiyuxing Zhang}
\orcid{0009-0002-9337-2278}
\affiliation{%
  \institution{Tsinghua University}
  \city{Beijing}
  \country{China}
}
\email{zxyx22@mails.tsinghua.edu.cn}

\author{Changhao Zhang}
\affiliation{%
  \institution{Ant Group}
  \city{Hangzhou}
  \country{China}
}
\email{eleven.zch@antgroup.com}

\author{Jian Liu}
\authornote{Co-corresponding authors.}
\affiliation{%
  \institution{Ant Group}
  \city{Hangzhou}
  \country{China}
}
\email{rex.lj@antgroup.com}

\author{Weiqiang Wang}
\affiliation{%
  \institution{Ant Group}
  \city{Hangzhou}
  \country{China}
}
\email{weiqiang.wwq@antgroup.com}

\author{Yuntao Wang}
\authornotemark[1]
\orcid{0000-0002-4249-8893}
\affiliation{%
  \institution{Tsinghua University}
  \city{Beijing}
  \country{China}
}
\email{yuntaowang@tsinghua.edu.cn}

\usepackage{booktabs}  % 引用三线表宏包
\usepackage{subcaption}
\usepackage{multirow}
\usepackage{array}
\usepackage{makecell}
\usepackage{threeparttable}
\usepackage{xcolor}

\newcolumntype{C}[1]{>{\centering\arraybackslash}p{#1}}

\begin{document}

\newcommand{\red}[1]{#1}
\newenvironment{redblock}
{}
{}

\title[AuthGlass: Benchmarking Voice Liveness Detection and Authentication ...]{AuthGlass: Benchmarking Voice Liveness Detection and Authentication on Smart Glasses via Comprehensive Acoustic Features}


\renewcommand{\shortauthors}{Xu et al.}

\begin{abstract}
 With the rapid advancement of smart glasses, voice interaction has been widely adopted due to its naturalness and convenience. However, its practical deployment is often undermined by vulnerability to spoofing attacks, while no public dataset currently exists for voice liveness detection and authentication in smart-glasses scenarios. To address this challenge, we first collect a multi-acoustic-modal dataset comprising 16-channel audio data from 42 subjects, along with corresponding attack samples covering two attack categories. Based on insights derived from this collected data, we propose AuthG-Live, a sound-field-based voice liveness detection method, and AuthG-Net, a multi-acoustic-modal authentication model. We further benchmark seven voice liveness detection methods and four authentication methods across diverse acoustic modalities. The results demonstrate that our proposed approach achieves state-of-the-art performance on four benchmark tasks, and extensive ablation studies validate the generalizability of our methods \red{under real-world constraints}. Finally, we release this dataset, termed AuthGlass, to facilitate future research on voice liveness detection and authentication for smart glasses.
\end{abstract}



\keywords{Do, Not, Use, This, Code, Put, the, Correct, Terms, for,
  Your, Paper}

\received{20 February 2007}
\received[revised]{12 March 2009}
\received[accepted]{5 June 2009}
\begin{teaserfigure}
  \includegraphics[width=\textwidth]{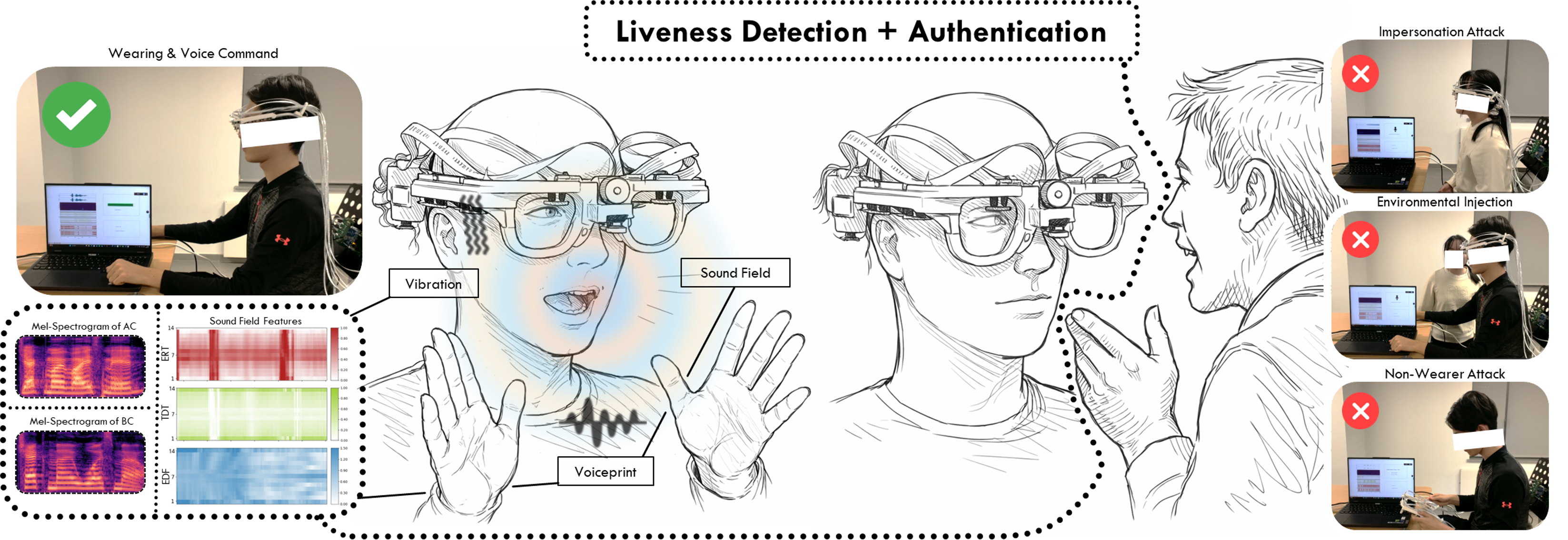}
  \caption{Based on insights derived from our collected multi-acoustic sensing data, we develop AuthG-Live, which performs liveness detection using sound-field features, and AuthG-Net, which conducts authentication by jointly leveraging voiceprint, vibration, and sound-field features. The proposed framework is generalizable to different input-channel configurations. In this figure, liveness detection and authentication are demonstrated using five air-conductive microphones (consistent with Ray-Ban Meta~\cite{MetaRayban2025}) mounted on the smart glasses and operating at a sampling rate of 16~kHz. As illustrated, the system successfully rejects three adversarial scenarios: \red{impersonation attack, environmental injection and non-wearer input}}
  \Description{}
  \label{fig:teaser}
\end{teaserfigure}

\maketitle

\section{Introduction}
Smart glasses have rapidly evolved from experimental prototypes to widely deployed wearable platforms, with commercial products such as Microsoft HoloLens~\cite{MicrosoftHololens} and Ray-Ban Meta Glasses~\cite{MetaRayban2025} gaining increasing adoption. In emerging application scenarios such as driving~\cite{he2015driving, chang2018design}, cooking~\cite{majil2022augmented, bouchard2020smart}, and industrial workflows~\cite{rajendran2020design}, users rely on smart glasses to access information and execute commands while their hands are occupied. As these devices become tightly integrated with users’ daily activities and increasingly handle sensitive personal or enterprise data, ensuring secure and reliable input mechanisms becomes critical~\cite{peng2016continuous}.

Existing authentication solutions for smart glasses span multiple paradigms. Commercial systems often rely on paired smartphones to provide \emph{device-level authentication}~\cite{MetaRayban2025, EchoFrames}, where access control is delegated to a trusted companion device. While effective for protecting device access, such approaches cannot distinguish between different users once the device is unlocked. Other user-level authentication methods, including password-based input~\cite{islam2018glasspass}, iris recognition~\cite{Glasspay, khade2021iris, li2017accurate}, and gaze-based techniques~\cite{fristrom2019free}, typically require explicit user interaction or specialized hardware, making them less suitable for seamless wearable use.

\red{Voice interaction has emerged as the dominant input modality on smart glasses due to its hands-free nature and minimal hardware requirements~\cite{RokidGlasses2025, MetaRayban2025}. It also enables implicit authentication by leveraging existing voice interfaces without requiring explicit user actions or additional sensors. However, this openness to ambient audio exposes voice-based systems to practical attacks, including command injection via loudspeakers, replayed recordings, and synthetic speech impersonation~\cite{zhang2016voicelive, yan2019catcher}, necessitating robust liveness detection and authentication mechanisms.}

Although voice authentication and liveness detection have been extensively studied on smartphones and other wearable devices~\cite{kersta1962voiceprint, voicepop, zhang2016voicelive, yan2019catcher, li2020vocalprint, feng2017continuous, blue20182ma, meng2022your, gao2021voice}, directly applying these approaches to smart glasses remains challenging. \red{A key limitation lies not only in the lack of datasets collected on smart-glasses platforms, but also in their unique sensing characteristics. Unlike handheld or ear-worn devices, smart glasses integrate multiple microphones along a curved frame around the head, capturing rich yet underexplored acoustic cues such as speech-induced vibrations and spatial sound-field patterns coupled with human anatomy. Existing methods, primarily designed for single-channel or regular-array configurations, do not account for these properties, making their effectiveness on smart glasses unclear.}

To address these challenges, we develop a DIY smart-glasses platform that captures 16-channel synchronized multi-acoustic-modal audio at 96~kHz, and build \emph{AuthGlass}, a dataset collected from 42 users with two representative attack sample sets. Based on this dataset, we analyze smart-glasses-specific acoustic characteristics and propose \emph{AuthG-Live}, a sound-field-based liveness detection method, and \emph{AuthG-Net}, a multi-modal authentication model that jointly leverages voiceprint, vibration, and spatial acoustic cues. We further benchmark seven liveness detection methods and four authentication methods across multiple modalities, establishing a systematic evaluation framework.

Our approach achieves over 96\% liveness detection accuracy and 97\% authentication accuracy, with strong generalization across attack types and robustness under cross-utterance settings. Extensive ablation studies demonstrate effectiveness under practical constraints and provide insights into microphone placement for smart glasses. To support future research, we open-source both the dataset and the smart-glasses platform implementation\footnote{We provide sample data and will release the full dataset and smart glasses prototype design files upon paper acceptance}.

Our key contributions are summarized as follows:
\begin{itemize}
\item We propose and open-source the AuthGlass dataset, which contains high-sampling-rate, 16-channel synchronized multi-acoustic-modal recordings from 42 participants, along with two attack scenarios. We also release the corresponding data capture platform.
\item We propose AuthG-Live, a voice liveness detection method using sound-field features that generalizes across different attacks, and AuthG-Net, a multi-acoustic-modal voice authentication model supporting utterance-independent authentication.
\item We benchmark seven voice liveness detection methods and four voice authentication methods across four tasks, demonstrating that our approach achieves state-of-the-art performance.
\item Through ablation studies, we show that our methods remain effective under limited acoustic channel settings, and provide design insights into microphone placement on smart glasses.
\end{itemize}

\section{Related Works}
This section summarize the related works, including voice interface on smart glasses (Section \ref{sec:relate1}), liveness detection and authentication on wearable devices (Section \ref{sec:relate2}), and benchmark for liveness detection and authentication on smart glasses (Section \ref{sec:relate3})

\subsection{Voice Interface on Smart Glasses}
\label{sec:relate1}

Voice user interfaces (VUIs) are a natural and intuitive interaction modality, widely adopted across domains such as smart homes \cite{iliev2022framework, torad2022voice, song2022investigation}, in-vehicle systems \cite{jung2020voice+, du2018voice}, and wearable devices \cite{chen2024enabling}. For smart glasses, VUIs rely only on lightweight microphones \cite{palermo2025advancements}, making them particularly suitable for resource-constrained form factors and thus widely deployed in commercial products such as Ray-Ban Meta \cite{MetaRayban2025}, Xreal One Pro \cite{Xreal2025}, and Rokid Glasses \cite{RokidGlasses2025}.
What's more, voice input supports hands-free command execution and conversational AI, especially with the emergence of large language models. For instance, Rokid enables voice-based payments, while Ray-Ban Meta integrates voice interaction with AI services. Prior work further highlights its benefits in accessibility (e.g., assisting visually impaired users \cite{AI-Powered4Speech2text, gamage2024ai, nithyaa2023eye, VisualImpair}) and in attention-critical scenarios such as medical and industrial settings \cite{zhang2025touchless, kim2016augmented}.

Despite these advantages, the security of voice interfaces on smart glasses remains underexplored. As devices increasingly handle sensitive personal data, they are exposed to realistic threats from ambient audio and adversarial inputs. Advances in speech synthesis and deepfake technologies \cite{kaur2023conventional, triantafyllopoulos2023overview, pham2025comprehensive} enable attacks such as replay and synthetic speech injection \cite{kamble2020advances, yi2023audio, ergunay2015vulnerability}, which are particularly concerning given the always-listening nature of smart glasses \cite{li2024sonicid, chan2015glass}. Prior studies show that authentication methods relying solely on acoustic features are vulnerable to replay \cite{yi2023audio, veesa2025deep} and synthetic attacks \cite{wenger2021hello, zuo2024advtts}, indicating limited robustness \cite{yan2022survey, li2023security}.

While voice liveness detection and authentication have been explored on other devices \cite{wang2019secure, zhang2022continuous, lu2020vocallock}, research targeting smart glasses remains limited. In particular, the lack of dedicated hardware platforms and standardized datasets hinders systematic evaluation. To address this gap, we develop and open-source a DIY smart glasses prototype that enables comprehensive multi-modal acoustic data collection, facilitating benchmarking for voice liveness detection and authentication in this domain.

\subsection{Liveness Detection and Authentication on Wearable Devices}
\label{sec:relate2}

Wearable devices such as earbuds and smartwatches are increasingly used as personal computing and interaction platforms \cite{seneviratne2017survey}. As they handle sensitive user data and enable command execution, securing access has become critical \cite{arias2015privacy}. Existing solutions can be broadly divided into \emph{liveness detection} and \emph{authentication}. Liveness detection distinguishes genuine human inputs from spoofed or adversarial signals, while authentication verifies user identity. Prior work has explored diverse sensing modalities, including audio \cite{ahmed2020void, he2024fast, SpeechDetection}, motion \cite{su2023gait, yang2015motionauth, chang2022vogue}, and physiological signals such as PPG \cite{zhao2020trueheart} and EMG \cite{shioji2018personal}. Among these, voice-based approaches \cite{ahmed2020void, voicelivenessdetection, gao2021voice} are particularly attractive due to their low hardware requirements, seamless integration with existing voice interfaces, and hands-free usability. For example, Huang et al. \cite{huang2025eve} exploit the consistency between air- and bone-conducted signals for liveness detection, while Gao et al. \cite{gao2021voice} utilize in-ear and out-ear acoustic features for authentication.

Compared to other wearables, smart glasses present distinct opportunities and challenges for voice-based security. Commercial devices \cite{MetaRayban2025, RokidGlasses2025, Xreal2025} typically integrate multiple microphones distributed along the frame, naturally enabling multi-channel audio capture with rich spatial and acoustic cues that can benefit liveness detection and authentication \cite{yang2023voshield, yan2019catcher}. 
Moreover, there is currently no unified hardware platform or publicly available dataset to systematically benchmark voice liveness detection and authentication in smart-glasses scenarios. 

To bridge this gap, we develop a multi-microphone DIY smart-glasses platform to collect multi-channel acoustic data, and build a benchmark with multi-acoustic modalities. Based on this, we further design \textit{AuthG-Live}, a sound-field-aware liveness detection method, and \textit{AuthG-Net}, a multi-acoustic-modality authentication model that jointly leverages voiceprint, vibration, and spatial acoustic features, enabling systematic evaluation and advancing robust voice-based security on smart glasses.

\subsection{Benchmark for Liveness Detection and Authentication on Smart Glasses}
\label{sec:relate3}

Modern smart glasses increasingly adopt multi-channel audio as a primary modality for voice interaction~\cite{RokidGlasses2025, MetaRayban2025, Xreal2025}, making robust liveness detection and authentication essential for security-sensitive use. While these problems have been extensively studied in automatic speaker verification (ASV), existing datasets are not tailored to smart-glasses scenarios.

Widely used datasets such as ASVspoof~\cite{yamagishi2021asvspoof}, VoxCeleb~\cite{nagrani2017voxceleb}, RedDots Replayed~\cite{kinnunen2017reddots}, and SITW~\cite{mclaren2016speakers} provide large-scale genuine and spoofed speech data for general ASV systems. However, they are primarily collected with single-channel or limited microphone setups and do not reflect the smart-glasses form factor. In particular, they fail to capture multi-channel acoustic characteristics induced by microphones distributed around the eyeglass frame, where spatial cues, channel-dependent energy patterns, and time delays are tightly coupled with device geometry. Complementary datasets such as REALMAN~\cite{yang2024realman} and the Massive Distributed Microphone Array Dataset~\cite{illinoisdatabankIDB-6216881} provide realistic multi-channel recordings using wearable or distributed microphone arrays. Nevertheless, they are not designed for ASV tasks and do not include replay or synthetic attacks, limiting their applicability for evaluating liveness detection and authentication.

Meanwhile, prior work on wearable or head-mounted devices~\cite{feng2017continuous, zhang2024safari, li2024boneauth} has explored voice-based liveness detection and authentication, but typically relies on custom hardware with heterogeneous microphone layouts and acquisition settings. The lack of publicly available datasets and standardized platforms hinders reproducibility and fair comparison, as summarized in Table~\ref{tab:wearable_comparison}.

To address these gaps, we introduce a unified multi-microphone smart-glasses platform and collect a comprehensive dataset from 42 users, covering both genuine speech and representative spoofing attacks. By releasing both the hardware design and dataset, our work provides, to the best of our knowledge, the first benchmark dedicated to voice liveness detection and authentication on smart glasses.

\begin{table}[t]
    \centering
    \caption{Comparison of voice-based liveness detection or authentication systems on wearable devices.}
    \label{tab:wearable_comparison}
    \small
    \setlength{\tabcolsep}{5pt}
    \renewcommand{\arraystretch}{1.15}
    \begin{tabular}{
        l
        >{\raggedright\arraybackslash}p{3cm}
        c
        c
        >{\raggedright\arraybackslash}p{3cm}
        c
    }
        \toprule
        \textbf{Name} &
        \textbf{Hardware} &
        \textbf{Sampling Rate} &
        \textbf{Channels} &
        \textbf{Modalities} &
        \textbf{Open-Source} \\
        \midrule

        VOGUE\cite{chang2022vogue} &
        Gyroscope embedded in existing device &
        44.1\,kHz / 50\,Hz &
        2 &
        Air-conducted microphone + gyroscope &
        $\times$ \\

        \midrule
        Vauth\cite{feng2017continuous} &
        IMU manually mounted on glasses nose pad &
        44.1\,kHz / 11\,kHz &
        2 &
        Air-conducted microphone + vibration sensor &
        $\times$ \\

        \midrule
        Eve Said Yes\cite{huang2025eve}&
        BC microphone module manually mounted &
        48\,kHz &
        2 &
        Air-conducted microphone + bone-conducted microphone &
        $\times$ \\

        \midrule
        Voshield\cite{yang2023voshield} &
        Microphone array module &
        16\,kHz &
        4 &
        Air-conducted microphones &
        $\times$ \\

        \midrule
        Ours &
        Multi-acoustic-modal microphone array on smart glasses frame &
        96\,kHz &
        16 &
        Multi air-conducted + bone-conducted microphones &
        $\checkmark$ \\

        \bottomrule
    \end{tabular}
\end{table}
\section{AuthGlass Dataset: Comprehensive Passive Acoustic Features for Liveness Detection and Authentication} 

This section introduce the AuthGlass Dataset that aims for two main objectives: \textbf{voice liveness detection} and \textbf{authentication} on smart glasses via passive acoustic sensing. Compared with existing datasets for authentication or liveness detection, our dataset offers four key advantages. \textbf{(a) Target on Smart Glasses}. To the best of our knowledge, AuthGlass is the first open-source acoustic dataset collected on a smart-glasses form factor for authentication and liveness detection. The dataset captures rich, multi-channel air- and bone-conducted signals from a glasses-mounted sensing platform that coupled with facial anatomy. \textbf{(b) Diverse Speakers and Spoofing Scenarios}. AuthGlass includes 42 participants, each providing 15 voice-command samples repeated at six volume levels, capturing substantial intra-speaker variability. The dataset further covers two spoofing categories with multiple attack scenarios, supporting systematic evaluation under diverse adversarial settings. \textbf{(c) 16-Channel Synchronized Multimodal Acoustic Capture}. The dataset captures both air- and bone-conducted acoustic signals, totaling 16 synchronized channels. This multimodal design enables the extraction of diverse acoustic features, including speech, vibration, and sound-field cues, and facilitates robust multimodal fusion for wearable authentication. \textbf{(d) High-Resolution Data}. Data are collected using a DIY smart-glasses prototype worn naturally by users, with all audio recorded at a 96 kHz sampling rate to preserve fine-grained acoustic details. Replay attacks are generated using high-quality playback devices, resulting in realistic and challenging attack conditions.

\red{While these advantages highlight the richness of AuthGlass, its design is driven by fundamental questions of motivation, threat models, and usable acoustic cues for smart-glasses-based authentication.
Accordingly, we first formalize liveness detection and authentication under explicit threat models and usage scenarios to clearly define the objectives of the dataset (Section~\ref{sec::chap3_motivation_and_objective}). We then identify key challenges introduced by the smart-glasses form factor (Section~\ref{sec::chap3_potential_acoustic}) and explore effective acoustic features across modalities to determine which signals should be collected (Section ~\ref{sec::chap3_validation_acoustic}). Finally, we integrate these insights to construct the AuthGlass dataset, including hardware design, data collection, and statistical analysis (Section~\ref{sec::chap3_multi-modal_acoustic_data_collection}).}

\begin{redblock}

\subsection{Motivation and Threat Model for AuthGlass Dataset Design}
\label{sec::chap3_motivation_and_objective}

In this section, we motivate voice authentication on smart glasses and systematically define the authentication pipeline and threat models, laying the foundation for the AuthGlass dataset design objectives.

\subsubsection{Motivation for Voice Liveness Detection and Authentication}

\begin{table}[t]
\centering
\caption{Comparison of user-level authentication modalities on smart glasses or head wearables.
\textbf{Task} indicates whether explicit user actions are required for authentication.
\textbf{Effort} denotes additional interaction beyond normal usage.
\textbf{Hands-free} indicates whether the modality supports fully hands-free interaction.
\textbf{Continuous Auth.} indicates whether authentication can be performed continuously without user input.
\textbf{Commodity HW} reflects common support on off-the-shelf smart glasses.
\textbf{Energy} is a qualitative estimate of runtime cost.
}
\label{tab:modality_compare}
\begin{tabular}{lcccccc}
\toprule
\textbf{Modality} & 
\textbf{Task} & 
\textbf{Effort} & 
\textbf{Hands-free} &
\textbf{Continuous Auth.} &
\textbf{Commodity HW} & 
\textbf{Energy}
 \\
\midrule

Touch (Password) \cite{islam2018glasspass} 
& $\checkmark$ & $\checkmark$ & $\times$ & No & $\checkmark$ & Low \\

Touch (Fingerprint) \cite{xu2025fingerglass}
& $\checkmark$ & $\times$ & $\times$ & Event-driven & $\times$ & Low \\

Gaze (Password) \cite{fristrom2019free} 
& $\checkmark$ & $\checkmark$ & $\checkmark$ & No & $\times$ & High \\

Gaze (Implicit Task) \cite{inoue2024user}
& $\times$ & $\times$ & $\checkmark$ & No & $\times$ & High \\

Iris recognition \cite{Glasspay, khade2021iris}
& $\checkmark$ & $\checkmark$ & $\checkmark$ & No & $\times$ & High  \\

Facial Biometrics \cite{lim2023c, li2024sonicid}
& $\times$ & $\times$ & $\checkmark$ & Continuous & $\times$ & High \\

Skull Biometrics \cite{schneegass2016skullconduct, shin2024skullid} 
& $\times$ & $\times$ & $\checkmark$ & Continuous & $\times$ & Medium  \\

Voice \cite{huang2025eve, zhang2021continuous}
& $\times$ & $\times$ & $\checkmark$ & Event-driven & $\checkmark$ & Low \\

\bottomrule
\end{tabular}
\end{table}

Voice interaction has emerged as the primary input modality on modern smart glasses due to its hands-free and natural interaction paradigm~\cite{MetaRayban2025, RokidGlasses2025, XiaomiSmartAudio, Xreal2025}. 
For example, users can issue voice commands to access information, control applications, or interact with AI assistants without interrupting their ongoing activities. 
Building authentication mechanisms directly on top of voice input therefore provides a seamless and intuitive solution that aligns with everyday usage.

Table~\ref{tab:modality_compare} compares voice-based authentication with existing modalities for smart glasses. 
Compared to touch- and gaze-based approaches, voice authentication does not require explicit user tasks or additional interaction effort, enabling a frictionless user experience. 
Unlike touch-based methods, it naturally supports hands-free interaction, which is essential for wearable devices. 
In contrast to vision-based biometrics (e.g., iris or facial contour recognition), voice authentication can be implemented using commodity microphones already available on current smart glasses, while incurring relatively low energy consumption. 
Although voice authentication is inherently event-driven rather than continuous, it aligns well with command-based interaction patterns, allowing authentication to be performed opportunistically during normal usage.

Despite these advantages, voice-based systems remain vulnerable to spoofing attacks (e.g., replay or synthesis) and environmental noise. 
Therefore, it is critical to systematically understand the role of voice authentication in practical systems. 
In particular, voice authentication is best suited for lightweight, event-driven verification without requiring additional user attention, while more secure fallback mechanisms (e.g., passwords on paired trusted devices) may still be needed for high-risk operations.

Motivated by these observations, we next formalize the authentication process for voice-based smart-glasses systems and define representative threat models and adversarial scenarios to guide the design of AuthGlass Dataset.

\subsubsection{Voice Authentication Process and Threat Model}
\begin{figure}
    \centering
    \includegraphics[width=1\linewidth]{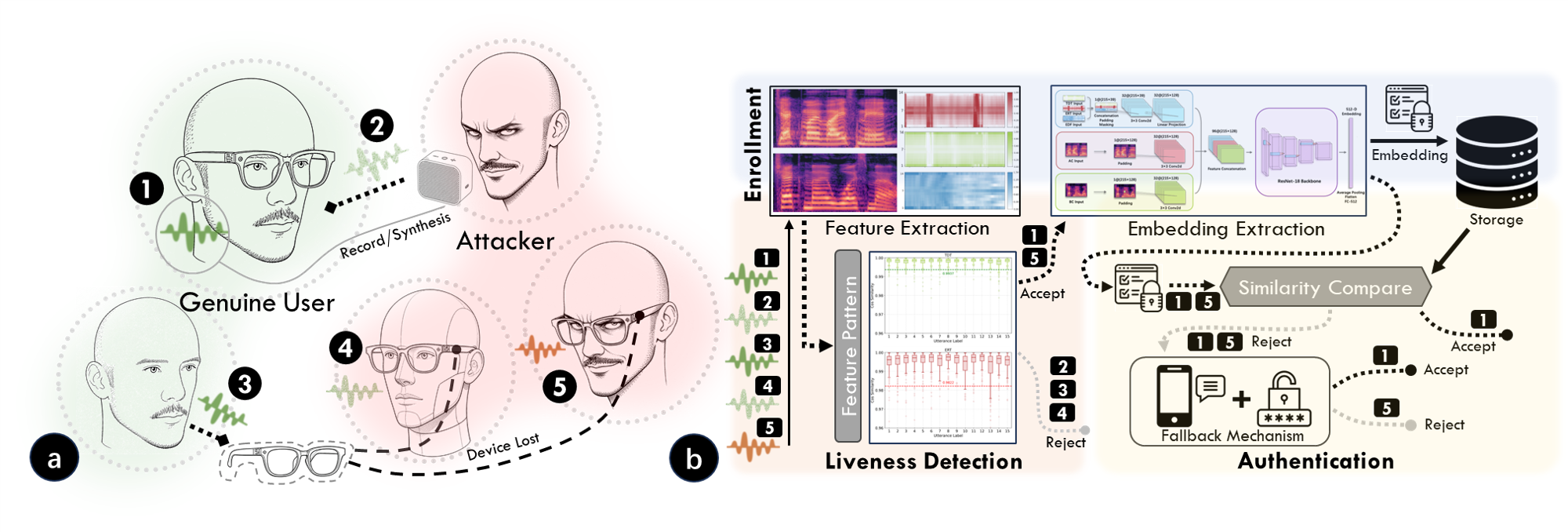}
    \caption{\red{Voice authentication pipeline and threat model. (a) Adversarial scenarios: (1) genuine user input, (2) environmental injection attacks, (3) non-wearer voice inputs, (4) advanced replay/spoofing attacks, and (5) device theft with impersonation. (b) System workflow: during enrollment, user voice embeddings are extracted and stored as templates. During operation, incoming speech first undergoes feature extraction and liveness detection, followed by embedding-based similarity comparison for authentication. Inputs from environmental injection attacks (2), non-wearer voice inputs (3), and advanced replay/spoofing attacks (4) are expected to be rejected at the liveness detection stage. In contrast, genuine user input (1) and device theft with impersonation (5) can pass liveness detection; the former is expected to be successfully authenticated, while the latter is rejected during the authentication stage. A fallback mechanism (e.g., smartphone-based verification) is triggered upon rejection to further prevent unauthorized access and to handle occasional false rejections of genuine users.}}
    \label{fig:authentication_model}
\end{figure}

\paragraph{Voice Authentication Workflow.}
We consider an event-driven voice authentication pipeline on smart glasses (Fig.~\ref{fig:authentication_model}b). Prior to deployment, each user completes a lightweight \textbf{enrollment} procedure by recording a small set of voice samples (e.g., passphrases), from which acoustic embeddings are extracted and stored as user-specific templates.

During operation, authentication is implicitly triggered by voice commands. Upon receiving input, the system first performs \textbf{liveness detection} to verify that the signal originates from a live speaker wearing the device, thereby rejecting replayed or externally injected audio. If successful, the system proceeds to \textbf{authentication}, where extracted features are matched against the enrolled template to verify identity. Commands are executed only if both stages succeed.

To account for practical variability (e.g., environmental noise, intra-user differences, or device conditions), voice authentication may occasionally result in false rejections. To maintain usability, we incorporate a fallback authentication mechanism (e.g., password input on a paired trusted device such as a smartphone), serving as a secondary authentication channel when voice-based verification fails. Compared to voice interaction, this fallback is less hands-free and attention-free, but provides a more robust and secure backup in rare failure cases.

\paragraph{Adversarial Scenarios and Threat Model.}
We model realistic adversarial scenarios based on a representative smart-glasses usage setting (e.g., industrial environments such as factory workers). We define four attack scenarios and explicitly specify the stage (liveness detection or authentication) each targets (Fig.~\ref{fig:authentication_model}a).

\textbf{(1) Environmental Injection Attacks (Liveness Failure Expected).}
An attacker injects malicious voice commands into the environment (e.g., via loudspeakers), attempting to trigger unintended operations. These commands may originate from text-to-speech (TTS), recorded user speech, or AI-generated voices. For example, an attacker replays a user’s recorded command near the victim to unlock restricted resources. Such attacks do not originate from a live wearer and should be rejected by liveness detection.

\textbf{(2) Non-wearer Voice Attacks (Liveness Failure Expected).}
Voice commands are issued when the legitimate user is not wearing the device (e.g., the glasses are placed on a table). Such external speech near the device may be captured. These inputs should also be rejected by liveness detection, as they do not correspond to a valid on-body speaking condition.

\textbf{(3) Advanced Spoofing with User Voice (Liveness Failure Expected).}
An attacker obtains recordings of the target user’s voice and attempts to reproduce them using advanced playback systems or humanoid simulators. While such attacks may mimic voice content, they cannot replicate the coupled acoustic characteristics of human speech production and anatomy. Therefore, they should be rejected by liveness detection.

\textbf{(4) Device Theft and Impersonation (Authentication Failure Expected).}
An attacker gains physical access to the smart glasses and wears the device (e.g., a co-worker attempting to access unauthorized data). In this case, the attacker produces live speech while wearing the device, and thus may pass liveness detection. However, the attacker’s voice characteristics differ from the legitimate user, and the system should reject the attempt during the authentication stage.

\paragraph{Attacker Capability Assumptions.}
We adopt the following assumptions regarding attacker capabilities:
\textbf{(a)} The attacker cannot compromise or manipulate the enrollment process or stored templates.
\textbf{(b)} The attacker may gain physical access to the device but cannot directly tamper with internal data streams or bypass the sensing pipeline.
\textbf{(c)} The attacker can obtain recordings of the target user’s voice (e.g., from public or environmental sources).
\textbf{(d)} The attacker cannot faithfully reproduce the user’s physiological acoustic characteristics, including vocal tract dynamics, bone-conducted signals, and head-related sound-field patterns.

\paragraph{Out-of-Scope and Limitations.}
We do not consider scenarios where an already authenticated device is later misused (e.g., after being left unattended). Due to the event-driven nature of voice authentication, identity is verified per command rather than continuously. Addressing this limitation requires complementary mechanisms (e.g., continuous authentication or wearing detection), which are beyond the scope of this work. More attack scenarios are discussed in Section \ref{sec::chap5_more_attacks}.

\end{redblock}

\subsection{Passive Acoustic Modality Selection and Challenges}
\label{sec::chap3_potential_acoustic}
In this section, we systematically analyze candidate passive acoustic features in terms of sensing feasibility, robustness, and device constraints, forming the basis for feature selection and the design of the AuthGlass dataset.

\begin{figure}
    \centering
    \includegraphics[width=0.8\linewidth]{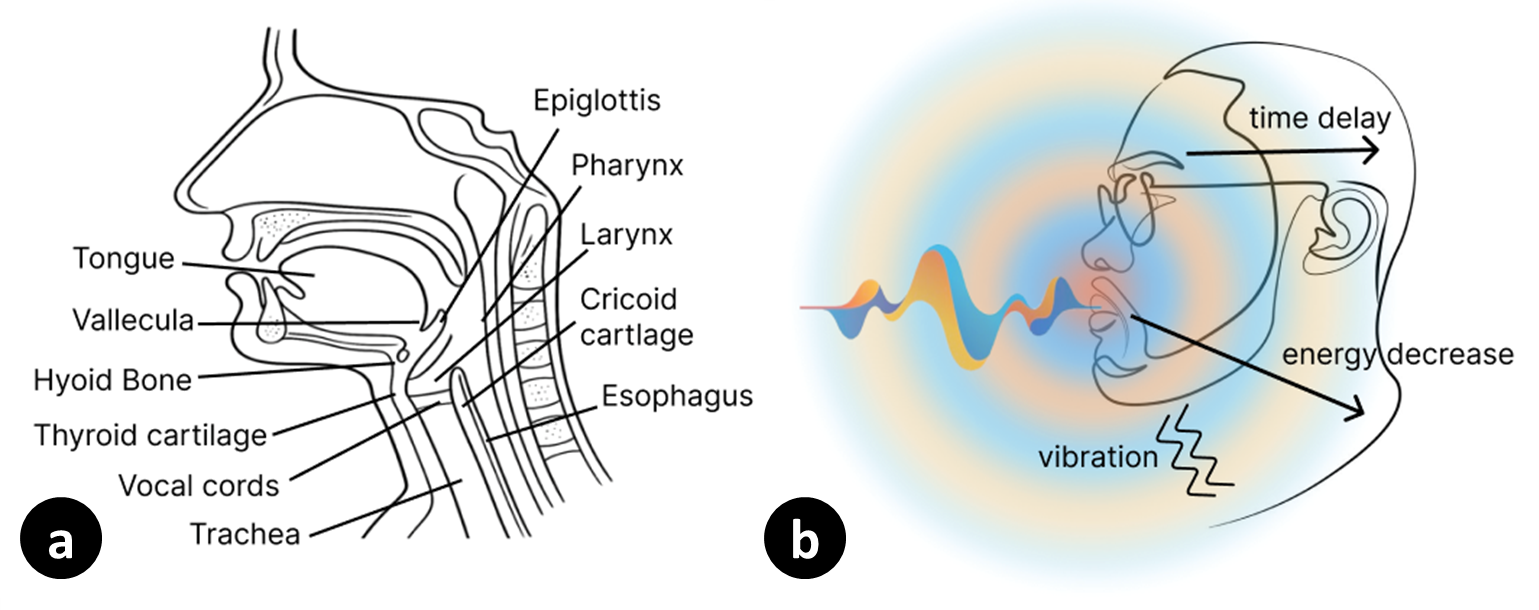}
    \caption{Human speech production system diagram. (a) Structure of the vocal tract, including the vocal cords, tongue, and other articulatory organs. (b) Acoustic features related to speech production, including vibrations conducted through skin and bones, as well as the spatial propagation effects of the sound field such as time delays and energy attenuation.}
    \label{fig:articulatory}
\end{figure}

\subsubsection{Human Articulatory System and Usable Passive Acoustic Modality}
\red{Voice-based authentication and liveness detection have been widely studied on conventional devices and various wearable platforms\cite{voicepop, zhang2016voicelive}. However, the specific acoustic modalities that can be effectively exploited in this context remain unclear.}
To address this question, we begin with the mechanism of human speech production (Fig.~\ref{fig:articulatory}). Airflow from the lungs excites the vocal folds, and the resulting sound source is subsequently shaped by the vocal tract into an acoustic signal with distinctive amplitude and spectral characteristics \cite{badin2006three}.
\red{
This process not only produces airborne speech but also induces coupled vibrations and propagation effects that are influenced by speaker-specific anatomical structures. Considering the sensing position and hardware constraints of smart glasses, a key question arises:} which types of speech-related cues can be effectively leveraged for liveness detection and authentication on such devices?

\subsubsection{Challenges on Leveraging Multi-Modal Acoustic Features}

Research on acoustic liveness detection and speaker authentication for wearables or smart devices \cite{ahmed2020void, huang2025eve, yan2019catcher, wang2019secure} has primarily explored four categories of features:
(i) voiceprint features extracted from airborne speech,
(ii) vibration-related features arising from speech transmission through human skin,
(iii) sound field features that characterize the spatial distribution of acoustic energy and phase, and
(iv) near-mouth acoustic cues observed in the close-to-microphone region.
\red{Inspired by these prior feature categories, we systematically examine their applicability to smart-glasses-based sensing, and analyze the feasibility and associated challenges of transferring these modalities under the constraints of the smart glasses form factor.}

First, prior studies have demonstrated that voiceprint features essential for speaker authentication can be reliably captured using a single airborne microphone \cite{ahmed2020void}. This modality is naturally compatible with smart glasses, as most off-the-shelf devices already incorporate airborne microphones for voice interaction. \red{Accordingly, a key question is where to place airborne microphones on smart glasses to effectively capture voiceprint features, which we investigate through pilot studies.}

Second, prior work \cite{li2024boneauth, huang2025eve} has shown that authentication systems relying solely on voiceprint features are vulnerable to spoofing attacks, motivating the use of speech-induced vibration signals as a complementary acoustic modality. Since smart glasses are in direct contact with the user’s head, such vibrations can be naturally sensed. Existing studies \cite{feng2017continuous, shi2020wearid} have explored the use of head-worn IMUs to capture these signals; however, their limited sampling rates (typically below 200 Hz) are insufficient to cover the frequency band of speech production (up to 8 kHz). In contrast, bone-conduction microphones have been shown to capture speech-induced vibrations across the speech frequency range \cite{huang2025eve}. \red{Prior systems such as SkullConduct \cite{schneegass2016skullconduct} and NasoVoice \cite{rekimoto2026nasovoce} also employ bone-conduction sensing on head-worn devices with different sensor placements.
Inspired by these observations, we conduct pilot studies to evaluate candidate placements of bone-conduction sensors on smart glasses, and analyze their effectiveness for capturing bone-conductive signals.}

Third, as an alternative approach to strengthening voiceprint-based authentication, sound field features that capture the spatial distribution of acoustic energy and phase have been shown to be effective \cite{yan2019catcher, yang2023voshield}. The eyeglass frame spans a relatively large area around the head, enabling multiple microphones to be distributed along the frame to sample the surrounding sound field. Indeed, many commercial smart glasses already integrate multiple microphones for audio capture and enhancement \cite{RokidGlasses2025, MetaRayban2025}, demonstrating the practical feasibility of sound field sensing.
\red{However, unlike prior work that typically assumes planar or matrix-structured microphone arrays, microphones on smart glasses are arranged along a curved, non-uniform geometry following the eyeglass frame. This structural difference challenges the applicability of existing sound-field analysis methods that rely on regular array layouts. Inspired by this observation, we conduct pilot studies to examine how such curved microphone configurations affect the extraction and usability of sound field features.}

Finally, prior studies have shown that close-to-mic acoustic features, such as plosive-related pop sounds \cite{voicepop} and airflow-induced signals \cite{wang2019secure}, can be leveraged for liveness detection and speaker authentication. However, these approaches typically rely on dedicated near-mouth microphones, and the use of such features has not been explored on eyeglass-mounted platforms. \red{However, these approaches typically rely on dedicated near-mouth microphones, and the use of such features has not been explored on eyeglass-mounted platforms. The proximity of the lower rim of the eyeglass frame to the mouth may enable the capture of similar close-to-mic acoustic cues.} Inspired by this observation, we investigate whether such features can be effectively captured using smart glasses.

To examine the usability of these features and \red{guide the design of the AuthGlass dataset}, we conduct a series of pilot studies to address the following questions:
\begin{itemize}
\item \textbf{Air-Conductive Feature (AC):} whether air-conduction microphones on smart glasses can reliably capture user-specific voiceprint information, and how their placement affects feature quality;
\item \textbf{Vibration or Bone-Conductive Feature (BC):} whether bone-conduction microphones can be effectively integrated into smart glasses to acquire vibration signals, and how different placements influence their reliability;
\item \textbf{Sound Field Feature (SF):} whether multiple air-conduction microphones on smart glasses can capture sound field information (e.g., spatial variations of acoustic energy and phase), and how such features can be effectively leveraged under non-uniform array configurations;
\item \textbf{Close-to-mic Features:} whether close-to-microphone features can be captured using sensors positioned near the mouth on smart glasses.
\end{itemize}

\subsection{Multi-Modal Acoustic Feature Exploration and Design}
\label{sec::chap3_validation_acoustic}
\red{To address the questions raised in Section~\ref{sec::chap3_potential_acoustic}, we conduct four pilot studies to systematically investigate the feasibility of different passive acoustic modalities on smart glasses. Based on these studies, we identify three effective feature categories—air-conducted (AC), bone-conducted (BC), and sound-field (SF) features.} We first built a proof-of-concept smart-glasses frame with microphones mounted at different positions to capture data from multiple acoustic modalities. The pilot study involved three participants, each producing six repetitions of randomly selected passphrases. After data collection, we analyze whether each modality can be used for liveness detection and authentication.

\begin{figure}
    \centering
    \includegraphics[width=1.0\linewidth]{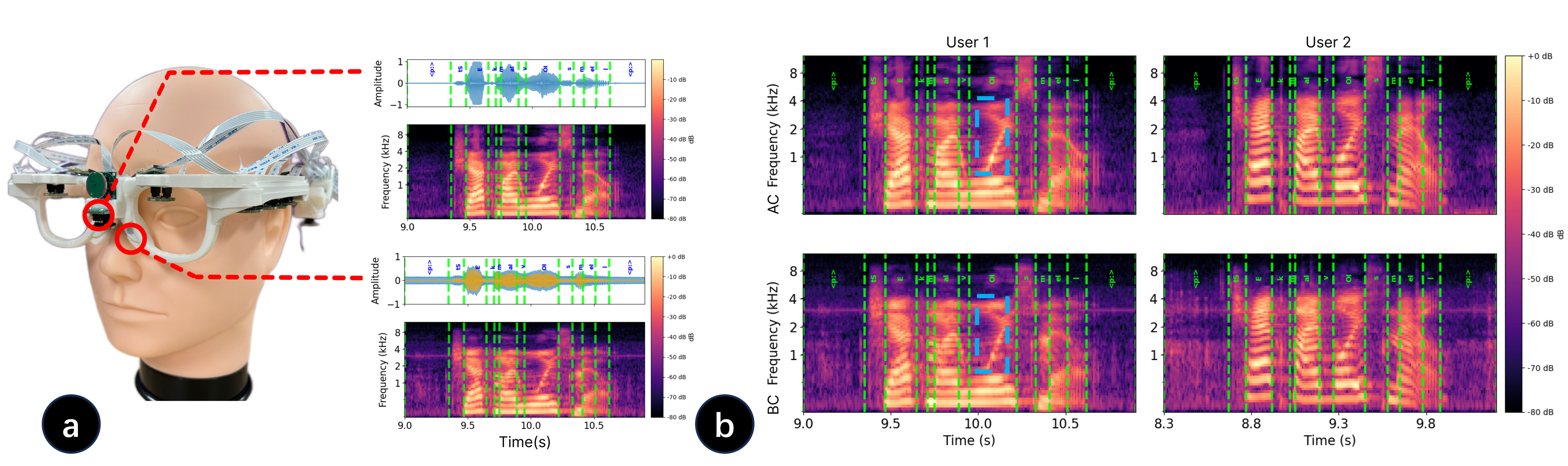}
    \caption{This figure illustrates the air-conductive (AC) and bone-conductive (BC) microphones used for feature selection. (a) The raw audio signals and corresponding Mel-spectrograms captured by AC and BC microphones when a user speaks the phrase “check my voicemail.” The utterance is segmented into phonemes. The results show that both AC and BC microphones capture rich time–frequency characteristics. (b) Comparison of air-conducted (AC) and bone-conducted (BC) signals from different users uttering “check my voicemail,” visualized using a consistent color scheme. As highlighted by the blue boxes, AC and BC signals from the same user exhibit subtle frequency-related differences, whereas signals captured by either AC or BC microphones show pronounced inter-user variability.}
    \label{fig:AC-BC}
\end{figure}

\subsubsection{Study 1: AC features}
To capture air-conducted (AC) features, we mounted an air-conductive microphone at the center of the glasses frame (Fig.~\ref{fig:AC-BC}a). The microphone records audio at a sampling rate of 96~kHz. From the raw recordings, we computed Mel-spectrograms and segmented them at the phoneme level using MAUS~\cite{Maus}. The signal energy is primarily concentrated in the 0--8~kHz range; therefore, we applied an 8~kHz low-pass cutoff. The results indicate that the centrally placed AC microphone captures rich voiceprint-related information, with each phoneme establishing a unique time--frequency pattern. We further compared Mel-spectrograms from two users uttering the same phrase. As shown in Fig.~\ref{fig:AC-BC}b, the AC signals exhibit clear time--frequency differences across users, suggesting that AC sensing on smart glasses can capture discriminative voiceprint features for authentication. \red{In contrast to the central placement, microphones along the frame capture weaker and less discriminative signals.}

\subsubsection{Study 2: BC features}
Similar to Study~1, we utilized a voice picking unit (VPU) to capture bone-conducted (BC) signals. As the VPU requires direct contact with the human body for data acquisition, we initially placed the sensor on the nose pad \red{\cite{rekimoto2026nasovoce}} and on the inner side of the glasses leg \red{\cite{schneegass2016skullconduct}}, forming contact with the nose and head, respectively. The results show that when placed on the inner side of the glasses leg, the captured signals are significantly weaker and contain noticeable spike noise. We attribute this to poor contact conditions and potential friction with hair. Based on these observations, we selected the nose pad as the placement location for BC sensing.

After determining the sensor placement, we used the VPU (96kHz) mounted on the nose pad of the smart glasses to capture bone-conducted data, as shown in Fig.~\ref{fig:AC-BC}a. We analyzed the recorded signals using Mel-spectrograms and segmented them at the phoneme level with MAUS\cite{Maus}. As shown in Fig.~\ref{fig:AC-BC}b, the VPU on smart glasses captures rich vibration data with time--frequency patterns similar to those of AC signals, while exhibiting subtle but consistent differences in the time--frequency domain, as highlighted by the blue boxes. We applied an 8~kHz cutoff to the BC signals and compared both AC and BC features across two users. Similar to AC signals, BC signals exhibit clear time--frequency differences across users, indicating that BC sensing provides complementary information that can enhance AC-based voiceprint authentication on smart glasses.

\begin{figure}
    \centering
    \includegraphics[width=1.0\linewidth]{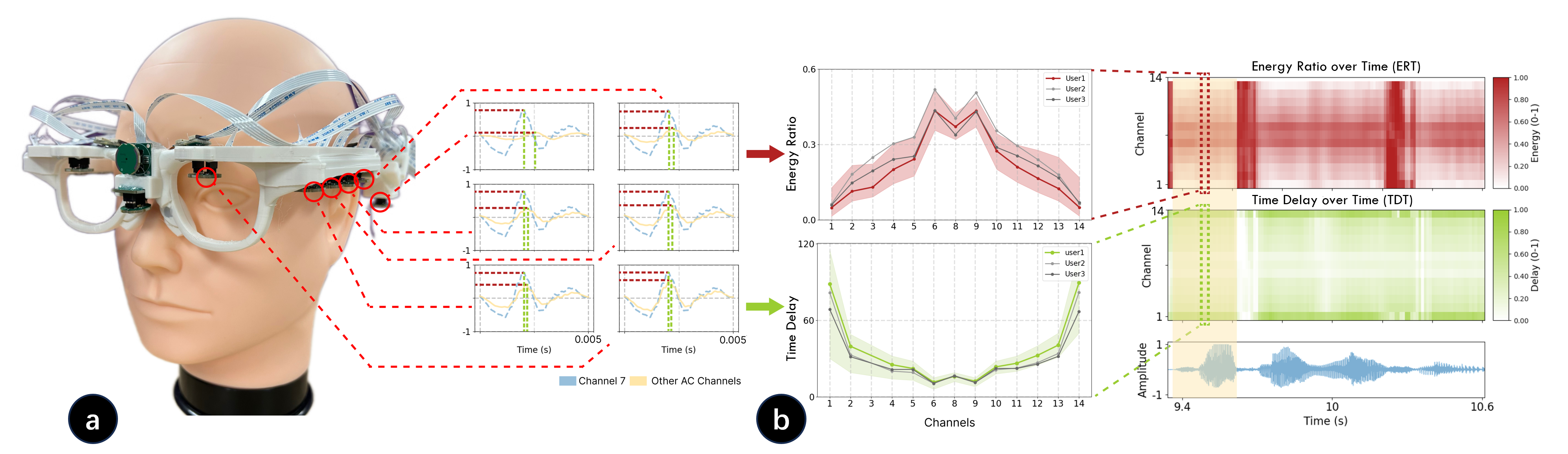}
    \caption{This figure illustrates the computation of Energy Ratio over Time (ERT) and Time Delay over Time (TDT), along with the spatial acoustic information they capture. (a) Energy attenuation (red dashed line) and time delay (green dashed line) between a selected AC channel and the central AC channel within a short time frame. Both energy attenuation and time delay vary across different AC channels. (b) Normalized energy ratio and time delay distributions across all time frames when a user speaks the phrase “check my voicemail”, forming the ERT and TDT. The left panel shows the averaged energy ratio and time delay distributions of different users, with inter-user variations highlighted by colored regions.}
    \label{fig:soundfield}
\end{figure}

\subsubsection{Study 3: SF features}
We further mounted multiple synchronized air-conducted (AC) microphones (96~kHz sampling rate) along the smart-glasses frame to capture sound field features during speech. As shown in Fig.~\ref{fig:soundfield}a, both energy attenuation and time delays between selected AC channels and the centrally placed AC microphone vary within a short time window. The centrally placed microphone captures the highest energy, while microphones farther from the mouth exhibit reduced energy. These observations indicate that sound-energy decay and phase shifts during acoustic propagation can be captured using spatially distributed AC microphones.

To further extract these features, we segmented the recordings into short time frames. For each frame, we measured the time delay between each selected channel and the central microphone using cross-correlation, and computed the corresponding energy attenuation, as shown in Fig.~\ref{fig:soundfield}b. To normalize feature distributions across channels, both the time-delay and energy-ratio values were scaled and passed through a hyperbolic tangent function, yielding values in the range $[-1, 1]$. \red{As the microphones are arranged in an array along the frame of the smart glasses, we organize them into a vector, resulting in a symmetrical pattern.} Stacking consecutive time frames reveals consistent temporal patterns in both energy attenuation and time delay during speech. We denote these features as Channel Energy Ratio over Time (ERT) and Channel Time Delay over Time (TDT) respectively. As shown in Fig.~\ref{fig:soundfield}b, averaging ERT and TDT along the time axis yields distinctive patterns across the three users, potentially related to individual head geometry.

\begin{figure}
    \centering
    \includegraphics[width=1.0\linewidth]{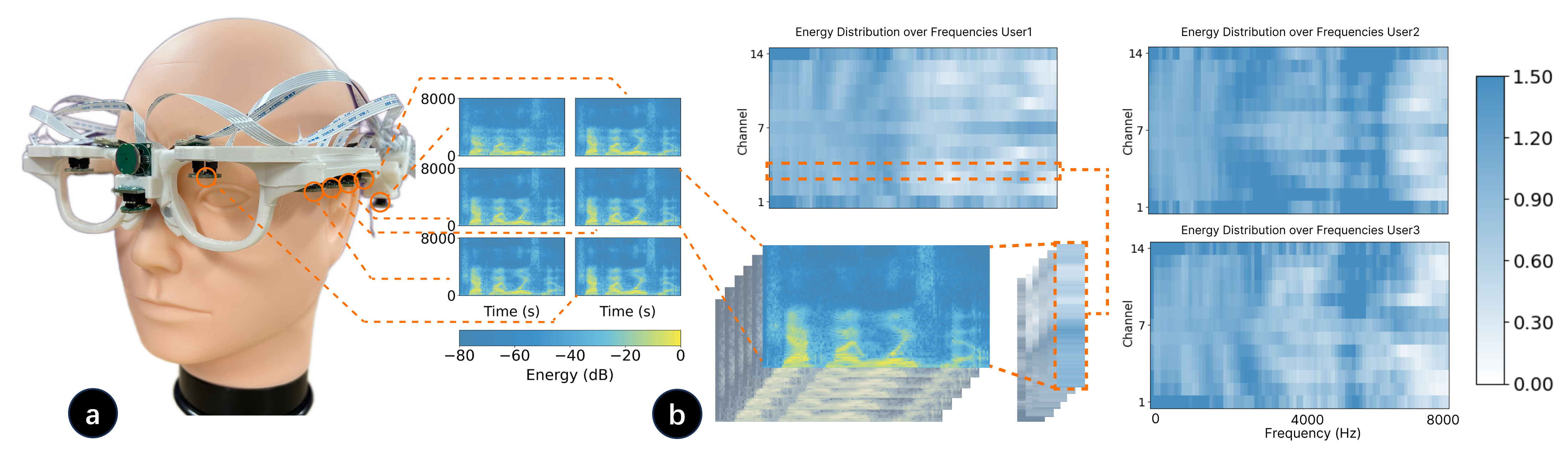}
    \caption{This figure illustrates the computation of Energy Distribution over Frequencies (EDF) and the spatial acoustic information it captures. (a) STFT of signals recorded by microphones at different positions on the glasses. (b) For each microphone, the STFT is averaged along the time axis to obtain the energy–frequency representation. On the right, EDFs for different users uttering the phrase “check my voicemail” are shown, highlighting inter-user variations.}
    \label{fig:soundfield_edf}
\end{figure}

In addition to time-domain features, we also examined sound-field characteristics in the frequency domain. Inspired by prior work on frequency-domain energy distributions~\cite{yan2019catcher,yang2023voshield}, we analyzed channel-wise energy attenuation across frequencies, as shown in Fig. \ref{fig:soundfield_edf} a. Specifically, we applied short-time Fourier transform (STFT) to the raw signals of each selected channel and averaged the resulting spectra along the time axis to obtain energy--frequency representations. We then computed the energy attenuation between each selected channel and the central microphone within specific frequency bands, and stacked the results across channels, as shown in Fig. \ref{fig:soundfield_edf}. The resulting frequency-domain energy distribution exhibits user-specific patterns and inter-user variability, suggesting its potential utility for authentication. We refer to this feature as Channel Energy Distribution over Frequencies (EDF).

\subsubsection{Study 4: Close-to-mic features}
To examine whether close-to-microphone features can be captured using sensors mounted on smart glasses, we placed an airflow sensor and microphones along the lower rim of the eyeglass frame, adjacent to the lenses, to capture airflow and plosive-related pop sounds, following prior works\cite{voicepop, wang2019secure}. However, we did not observe discernible close-to-microphone features under this configuration.

\subsection{AuthGlass Dataset}
\label{sec:dataset}

\subsubsection{Design Overview}
\begin{table}[t]
\centering
\caption{Summary of dataset classes and configurations}
\label{tab:dataset_category_summary}
\begin{tabular}{
>{\raggedright\arraybackslash}p{3cm}
l
>{\small\raggedright\arraybackslash}p{4cm}
l
c
}
\toprule
\textbf{Category} & \textbf{Name} & \textbf{Description} & \textbf{Subjects} & \textbf{Utterances / Subject} \\
\midrule
\textbf{Genuine User Data} & Positive Sample & The genuine user data collected and segmented in section \ref{sec:dataset} & 42 & 15 (6 repeats) \\
\midrule
\textbf{Wearer-based Attack Data} & Attack Setting 1 (Set 1) & The replay data generated by simulated wearers in section \ref{sec:dataset} & 42 & 15 (6 repeats) \\
\midrule
\multirow{6}{*}{\parbox{3 cm}{\raggedright\textbf{Environment-based Attack Data}}}
       & Attack Setting 2 (Set 2) & 100 cm in front of the user. & 42 & 15 (6 repeats) \\
       & Attack Setting 3 (Set 3) & 100 cm on right side. & 42 & 15 (6 repeats) \\
       & Attack Setting 4 (Set 4) & 100 cm in back of the user. & 42 & 15 (6 repeats) \\
       & Attack Setting 5 (Set 5) & 100 cm on left side. & 42 & 15 (6 repeats) \\
       & Attack Setting 6 (Set 6) & 50 cm in front of the user. & 42 & 15 (6 repeats) \\
       & Attack Setting 7 (Set 7) & 25 cm in front of the user. & 42 & 15 (6 repeats) \\
\bottomrule
\end{tabular}
\end{table}

\paragraph{Dataset Design under Threat Models.}
\begin{redblock}
    
Building upon the threat models and design objectives discussed in Section~\ref{sec::chap3_motivation_and_objective}, the dataset is constructed to systematically cover four representative attack categories, including \textit{(1) Environmental Injection Attacks}, \textit{(2) Non-wearer Voice Attacks}, \textit{(3) Advanced Spoofing with User Voice}, and \textit{(4) Device Theft and Impersonation}. For attack scenarios (1) and (2), genuine voice inputs collected from the legitimate wearer are treated as positive samples, while adversarial samples are generated by replaying audio through external speakers in the environment, simulating injected or non-wearer speech.  
For scenario (3), adversarial samples are collected using humanoid simulators that reproduce the target user’s voice, representing more advanced spoofing attempts.  
For scenario (4), data are collected from multiple real users, where each user may act as an attacker attempting to impersonate another enrolled user while wearing the device.  

Accordingly, the dataset is organized into three major data categories: \textit{(i) genuine user data}, \textit{(ii) wearer-based attack data}, and \textit{(iii) environment-based attack data}. Table~\ref{tab:dataset_category_summary} summarizes the detailed data composition and their correspondence to different attack scenarios. This threat-driven design enables systematic evaluation of both liveness detection and authentication under diverse and realistic adversarial conditions, as illustrated in Section \ref{sec:chap4}.

\begin{table}[t]
    \centering
    \caption{Summary of modalities and channel configurations}
    \label{tab:dataset_modality_channel}
    \begin{tabular}{
        l
        >{\raggedright\arraybackslash}p{4cm}
        >{\small\raggedright\arraybackslash}p{6.5cm}
    }
        \toprule
        \textbf{Modality} & \textbf{Channel} & \textbf{Description} \\
        \midrule
        \textbf{AC} & Channel 7 &
        AC channel mounted centrally and facing downward. \\

        \midrule
       \textbf{BC} & Higher-energy channel (15, 16) &
        Bone-conduction channel with higher signal energy selected as the representative BC input. \\

        \midrule
        \textbf{SF} & Channels [1--6], [8--14] &
        All remaining AC channels that can function as complementary channels in comparison with the central AC channel. \\
        \bottomrule
    \end{tabular}
\end{table}

\paragraph{Multi-Modal Acoustic Data Collection.}
\label{sec::chap3_multi-modal_acoustic_data_collection}
Based on the validation of acoustic modalities in Section~\ref{sec::chap3_validation_acoustic}, the dataset captures three complementary acoustic modalities, including \textit{air-conducted (AC)}, \textit{bone-conducted (BC)}, and \textit{sound-field (SF)} signals. These modalities correspond to different physical propagation paths of speech, including airborne transmission, vibration through human tissues, and spatial acoustic distribution around the head. To enable synchronized acquisition of these modalities, we implement a DIY smart-glasses platform (Section~\ref{sec::chap3_hardware_implementation}) equipped with 16-channels microphone array consisting of both air-conduction and bone-conduction sensors.
Table~\ref{tab:dataset_modality_channel} summarizes the definition of each modality and its corresponding sensing channels. 
\end{redblock}

\subsubsection{Hardware Implementation}
\label{sec::chap3_hardware_implementation}

\begin{figure}
    \centering
    \includegraphics[width=1.0\linewidth]{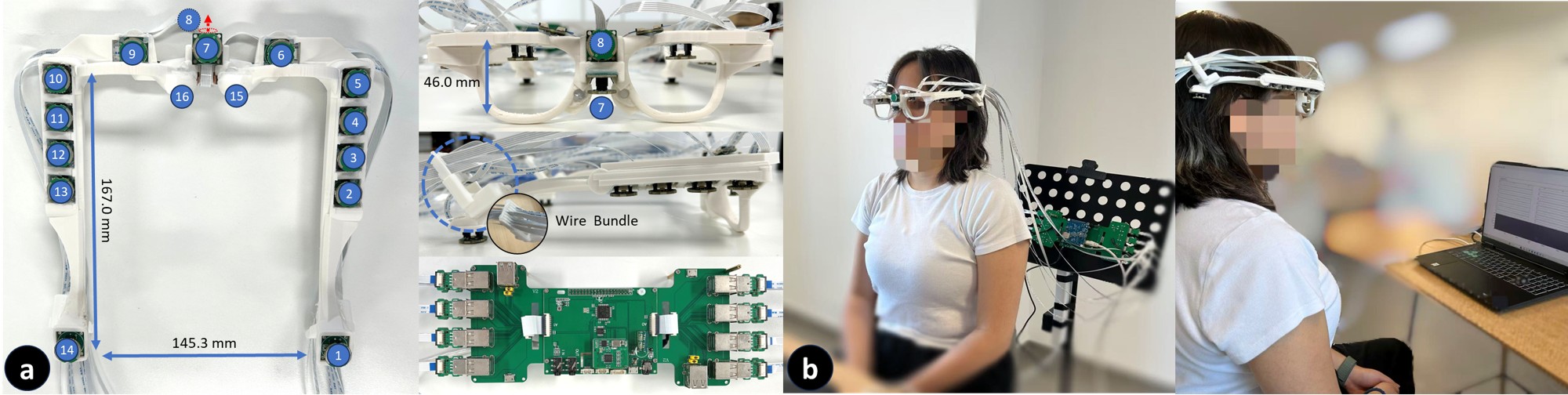}
    \caption{(a) Structural and hardware overview of the prototype. The system comprises 14 air-conduction microphones and 2 bone-conduction microphones. Each microphone is connected to the data collection board via flexible flat cables, and the temple is designed with a wire bundle, which can reduce crosstalk between channels. b) Illustration of a user wearing the smart glasses prototype.}
    \label{fig:data_collection}
\end{figure}

To support multi-modal data acquisition, we implement a custom DIY smart-glasses prototype, as illustrated in Fig.~\ref{fig:data_collection}a. The prototype integrates 14 omnidirectional air-conduction (AC) microphones (Infineon IM73D122V01~\cite{Infineon}) and two bone-conduction (BC) voice picking units (VPU14AA01~\cite{Sonion}), with overall dimensions of 167.0~mm (length) and 145.3~mm (width).

To enable spatial acoustic sampling for sound-field analysis, the AC microphones are distributed across the eyeglass frame: five on each temple, one above each lens, and two at the center (facing downward and forward, respectively), providing diverse spatial coverage.

Guided by pilot study insights, one BC sensor is embedded in each nose pad to capture vibration signals while mitigating degradation caused by asymmetric skin contact. All sensors are connected via bundled flexible flat cables (FFCs) to a data collection board~\cite{Collection_board}, enabling synchronized acquisition of 16 channels at 96~kHz. The recorded data are streamed to a PC via TCP for subsequent processing.

\subsubsection{Data Collection}
\begin{figure}
    \centering
    \includegraphics[width=1.0\linewidth]{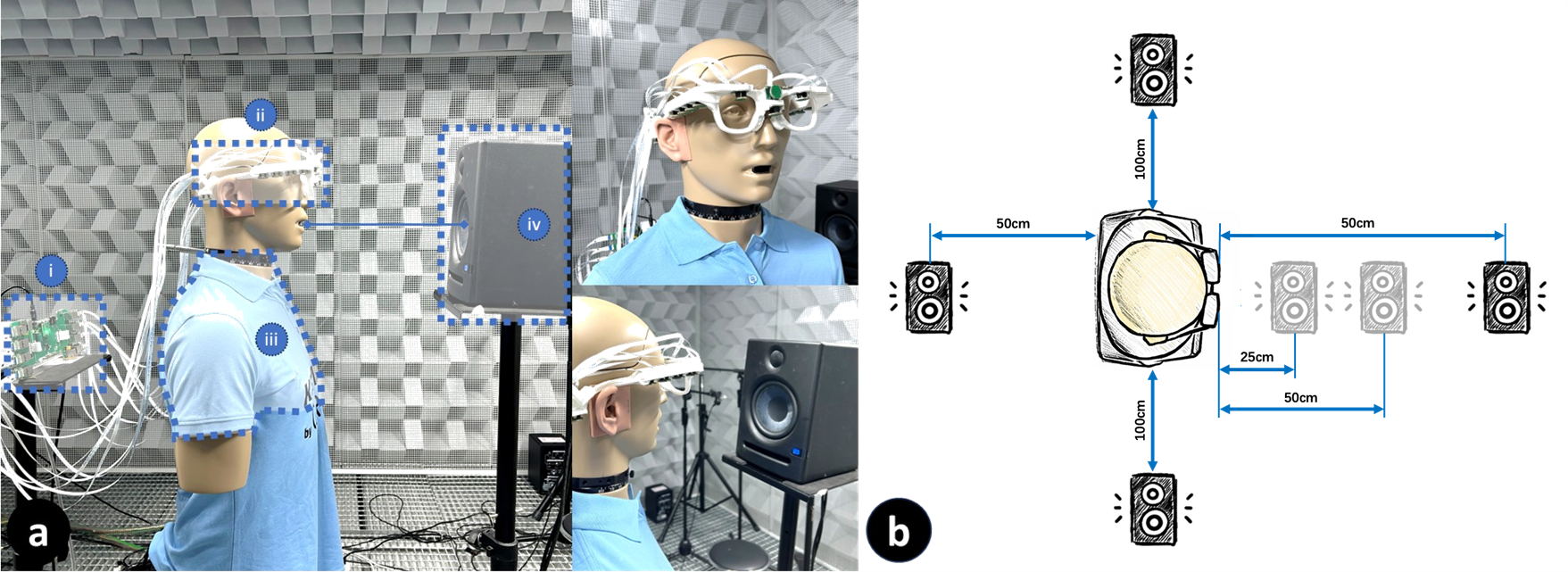}
    \caption{(a) The environment setting of the voice attack. i is the data collection board and ii is the smart glasses frame. All microphones are connected via bundled flexible flat cables to a data collection board. iii is the torso–mouth simulator and iv is the loudspeaker. In the experiment, two types of replay sample were applied: loudspeaker replay and a torso–mouth simulator replay which replicates the exhalation process and torso vibrations during speaking. (b) In the replay attack, the loudspeaker was placed at different positions and distances from the smart glasses to simulate various attack situations. }
    \label{fig:attack_collection}
\end{figure}

We then collected data using our smart-glasses prototype for positive sample and two other attack categories, described in Table~\ref{tab:dataset_category_summary}. The study protocol was approved by the Institutional Review Board (IRB). All data are anonymized for dataset publication.

\paragraph{Genuine User Data (Positive Sample).}
We recruited 42 participants (14 female and 28 male), aged 18–42 \red{(mean $24.6 \pm 4.3$)}, from a university campus. \red{Detailed participant demographics are provided in Appendix~\ref{app:demographic}.}. All participants provided informed consent, permitting the use and publication of their anonymized data for research purposes. Data collection was conducted in a quiet conference room to minimize environmental noise. Participants were asked to wear our glasses prototype and were seated in a natural posture and instructed to avoid excessive head or body movements during recording. As illustrated in Fig.~\ref{fig:data_collection}(b), each participant uttered a predefined set (a total of 15) of voice passphrases (see Appendix~\ref{sec:passphrase_set}) at three distinct volume levels, ranging from soft speech (e.g., speaking to a nearby listener) to loud speech (e.g., projecting to a listener approximately 5 meters away). Each passphrase–volume combination was repeated twice, resulting in six repetitions per sentence. Short rest intervals were provided between passphrases to reduce vocal fatigue. All utterances were captured by the glasses-mounted microphones and manually reviewed to confirm successful recording. The audio was subsequently segmented using MAUS~\cite{Maus}, with the resulting segments manually verified for accuracy.

\paragraph{Wearer-based Attack Data (Attack Setting 1).}
To collect wearer-based attack data, we conduct experiments in a sound isolation room, as shown in Fig.~\ref{fig:attack_collection}a. \red{The smart glasses are mounted on a GRAS 45BC KEMAR torso-mouth simulator~\cite{KEMAR}, which replays genuine utterances through its built-in speaker. The playback volume is manually calibrated to match the sound pressure levels observed during genuine data collection. All genuine samples are replayed via the simulator’s mouth and recorded by the glasses.}

\paragraph{Environment-based Attack Data (Attack Setting 2--7).}
To collect environment-based attack data, we use the torso-mouth simulator as a mannequin to emulate a stationary user. \red{Genuine utterances are replayed using a high-quality loudspeaker (PreSonus Eris E5~\cite{Eris}) positioned at the same height as the mannequin, with playback volume manually calibrated.} To model diverse attack source locations, the speaker is placed at six distinct positions around the mannequin, corresponding to different directions and distances (Fig.~\ref{fig:attack_collection}b). For each position, all genuine samples are replayed and recorded by the glasses.

\subsubsection{Statistics Analysis}
After data collection, we obtained AuthGlass datasets that consists of three categories of data. Each sample consisting of 16 synchronized audio channels at 96 kHz. For \textit{Genuine User Data}, the dataset contains a total of $42 \times 6 \times 15 = 3{,}780$ utterances, corresponding to 15 passphrases (length from 2 to 5 words) spoken at three volume levels by each participant. For \textit{Wearer-based Attack Data (Attack Setting 1)}, we collected 3,780 attack samples, matching the size and content of the genuine dataset. For \textit{Environment-based Attack Data (Attack Setting 2--7)}, each genuine utterance was replayed and recorded at six distinct speaker locations, yielding $3{,}780 \times 6 = 22{,}680$ attack recordings. All recordings were segmented using MAUS~\cite{Maus} and temporally aligned and segmented, followed by manual verification of alignment and segmentation quality. 

\red{
While the dataset provides comprehensive multi-modal acoustic recordings, we note that the participant pool is still relatively limited in size and demographic diversity (e.g., primarily recruited from a university population), which may not fully capture broader variations in speech characteristics. A more detailed discussion of these limitations is provided in Section~\ref{sec::chap5_limit_dataset}.
The complete dataset, including raw multichannel recordings, segmentation annotations, and the design files of the smart-glasses prototype, has been publicly released, enabling future work to reproduce, extend, and enrich the dataset with more diverse participants and scenarios.
}
\section{Benchmarking Voice Liveness Detection and Authentication on Smart  Glasses}
\label{sec:chap4}
\red{In this section, we first design a set of benchmark tasks based on the collected data to evaluate voice liveness detection and authentication on smart glasses (Section~\ref{sec:benchmark_tasks_design}).} Then, we select seven liveness detection methods and three authentication methods to cover a range of evaluation scenarios (Section~\ref{sec:method_selection}). We further present our proposed lightweight liveness detection and authentication system (Section~\ref{sec:our_method}). Finally, we report experimental results across all benchmark tasks, showing that our method achieves state-of-the-art performance (Section~\ref{sec:benchmark_result}) with conclusion (Section~\ref{sec:authentication_conclusion})

\subsection{Benchmark Tasks Design}
\label{sec:benchmark_tasks_design}
We define four benchmark tasks (two for liveness detection and two for authentication) with task description and evaluation metrics, following the authentication and threat model in section \ref{sec::chap3_motivation_and_objective}.

\paragraph{Liveness Detection}
\red{aims to determine whether the input audio originates from a genuine glasses wearer. Ideally, it should reject all inputs that do not correspond to a live user wearing the device. In this work, we focus on three representative adversarial scenarios introduced in Section~\ref{sec::chap3_motivation_and_objective}, including \textit{Environmental Injection Attacks}, \textit{Non-wearer Voice Attacks}, and \textit{Advanced Spoofing with User Voice}. These scenarios are instantiated in our dataset through \textit{wearer-based attack data} and \textit{environment-based attack data} (Table~\ref{tab:dataset_category_summary})}. In addition, liveness detection approaches that rely on identifying specific attack devices or synthesized signals may fail to generalize to unseen attacks. This limitation is particularly critical in real-world scenarios, where attacks may originate from arbitrary and previously unknown devices. To investigate different methods' performance on this issue, we further design a task to examine the generalization of Liveness Detection to unseen attacks. \red{For each task, we use accuracy at the EER operating point, i.e., $1 - \mathrm{EER}$ as evaluation metric.}

\subsubsection{Task 1: Liveness Detection Accuracy}
\begin{redblock}
(a) Data Split.
We construct seven evaluation sets by pairing positive samples with attack samples from each attack type (Attack Setting 1–7), where each set contains an equal number of positive and attack samples with one-to-one correspondence. A subject-independent split is adopted, such that samples from a subset of subjects are used for training and the remaining subjects for testing, with no subject overlap. Positive and attack samples are labeled as 0 and 1, respectively.

(b) Method.
For each split, the liveness detection model is trained on the training set and evaluated on the test set. The trained model predicts the labels of both positive and attack samples in the test data. Performance is measured using overall classification accuracy, and results are averaged across multiple splits.
\end{redblock}

\subsubsection{Task 2: Generalization of Liveness Detection to Unseen Attacks}
\begin{redblock}
(a) Data Split.
To evaluate cross-attack generalization, we construct seven datasets by pairing positive samples with attack samples from each attack type (Attack Setting 1–7), following the same protocol as in Task 1. A subject-independent split is adopted, where samples from a subset of subjects are used for training and the remaining subjects for testing. The same subject split is consistently applied across all datasets to ensure fair comparison.

(b) Method.
For each split, the model is trained on one dataset (corresponding to a specific attack type) using the training subjects, and evaluated on the test portions of the remaining datasets (corresponding to unseen attack types) using the test subjects. For example, the model can be trained on $(\text{Positive}, \text{Set 1})$ samples from the training subjects, and then tested on $(\text{Positive}, \text{Set 2})$ to $(\text{Positive}, \text{Set 7})$ samples from the test subjects. The model predicts labels for both positive and attack samples, and performance is measured by classification accuracy. The final result is obtained by averaging accuracy across all training–testing combinations for each split.
\end{redblock}

\paragraph{Authentication} 
aims to distinguish voice inputs from different genuine users by accepting the enrolled user while rejecting all others. Accordingly, we first define a primary authentication task to evaluate the baseline performance of different methods.
In practical scenarios, an authentication system should support fast enrollment with minimal user effort, as requiring repeated utterances for each new voice command is impractical. To reflect this constraint, we further design a task to evaluate authentication performance for cross-utterance scenarios.

\subsubsection{Task 3: Authentication Accuracy}
\label{sec::benchmark_task3}
\begin{redblock}
(a) Data Split.
To evaluate authentication performance, we adopt a subject-independent split, where samples from a subset of subjects are used for training and the remaining subjects for testing. Only positive samples are used, as the task focuses on distinguishing among genuine users. Each sample is labeled with a subject ID (1–42) and an utterance ID (1–15).

(b) Method.
For each split, the authentication model is trained using the training set with corresponding subject and utterance labels. During testing, one sample per utterance is randomly selected from the test subjects to form the enrollment set, resulting in 15 enrollment samples per subject. This selection is kept consistent across all methods. The remaining samples are used as the test set. The model predicts the subject identity for each test sample, and performance is measured using classification accuracy.
\end{redblock}

\subsubsection{Task 4: Cross-utterance Authentication Performance}
\begin{redblock}
(a) Data Split.
To evaluate cross-utterance authentication performance, we adopt the same subject-independent split as in Task 3. Only positive samples are used, with each sample annotated by a subject ID (1–42) and an utterance ID (1–15).

(b) Method.
For each split, the authentication model is trained on the training set with corresponding subject and utterance labels. During testing, we simulate a cross-utterance enrollment scenario by selecting a subset of utterances per subject to form the enrollment set. Specifically, the 15 utterances are partitioned into multiple non-overlapping groups, and each group is used as enrollment while the remaining utterances form the test set. The selection of utterance groups is fixed and consistent across all methods to ensure fair comparison. 
For each configuration, the model predicts the subject identity of test samples, and authentication accuracy is computed. The final performance is obtained by averaging results across all configurations for each split.
\end{redblock}

\subsection{Existing Method and Adjustments to Multi-channel Acoustic Data}
\label{sec:method_selection}
We selected seven representative methods for voice liveness detection and four methods for authentication, spanning AC, BC, and SF acoustic modalities. We further modified these methods to accommodate multiple features, aligning their input formats for systematic comparison.

\begin{table}[t]
    \centering
    \caption{Comparison of Voice Liveness Detection Methods}
    \label{tab:liveness_comparison}
    \begin{tabular}{
        p{3.0cm}
        C{0.8cm}
        C{0.8cm}
        C{0.8cm}
        p{4.2cm}
        p{3.0cm}
    }
        \toprule
        \textbf{Method} & \textbf{AC} & \textbf{BC} & \textbf{SF} &
        \textbf{Feature} & \textbf{Model / Method} \\
        \midrule
        VOID \cite{ahmed2020void}                 & \checkmark & - & - & Manually Defined Features & Traditional ML \\
        He et al.(2024) \cite{he2024fast} & \checkmark & - & - & Manually Defined Features & CNN \\
        Boneauth \cite{li2024boneauth}             & \checkmark & \checkmark & - & MFCC & Cosine Similarity \\
        Eve Said Yes \cite{huang2025eve}         & \checkmark & \checkmark & - & STFT with pooling & Temporal Similarity \\
        Li et al.(2021) \cite{li2021robust}   & \checkmark & - & \checkmark & STFT with phase & CNN \\
        Cafield   \cite{yan2019catcher}           & - & - & \checkmark & Fieldprint & GMM \\
        Voshield \cite{yang2023voshield}             & - & - & \checkmark & Sound Field Dynamics & CNN \\
        \midrule
        Ours                 & - & - & \checkmark & TDT and ERT & Cosine Similarity \\
        \bottomrule
    \end{tabular}
\end{table}

\begin{table}[t]
    \centering
    \caption{Comparison of Voice Authentication Methods}
    \label{tab:auth_comparison}
    \begin{threeparttable}
    \begin{tabular}{
        p{3.0cm}
        C{0.8cm}
        C{0.8cm}
        C{0.8cm}
        p{4.2cm}
        p{3.0cm}
    }
        \toprule
        \textbf{Method} & \textbf{AC} & \textbf{BC} & \textbf{SF} &
        \textbf{Feature} & \textbf{Model / Method} \\
        \midrule
        Park et al. (2025) \cite{park2025toward} & \checkmark & - & - & MFCC & LSTM \\
        Eve Said Yes \cite{huang2025eve}            & - & \checkmark & - & CQT & CNN with AL \tnote{1}\\
        Cafield  \cite{yan2019catcher}  & - & - & \checkmark & Fieldprint & GMM \\
        Voshield\tnote{2}  \cite{yang2023voshield}                         & - & - & \checkmark & Sound Field Dynamics & CNN \\
        \midrule
        Ours                     & \checkmark & \checkmark & \checkmark & MFCC, TDT, ERT and EDF & CNN with AL \tnote{1} \\
        \bottomrule
    \end{tabular}
    \begin{tablenotes}
        \footnotesize
        \item[1] indicates methods employing adversarial learning.
        \item[2] the method originally do not support authentication.
    \end{tablenotes}
    \end{threeparttable}
\end{table}

\paragraph{Voice Liveness Detection Methods.}
We selected seven representative voice liveness detection methods, as summarized in Table~\ref{tab:liveness_comparison}. These methods span AC, BC, and SF acoustic modalities and employ diverse feature representations and detection models. To enable a fair comparison, we align all methods in terms of input modalities. \red{Methods and our corresponding modifications are introduced in Appendix \ref{app:method_selection}.}

\paragraph{Voice Authentication Methods.}
We selected four voice authentication methods, as summarized in Table~\ref{tab:auth_comparison}. Similar to the liveness detection methods, all authentication methods are modified to accommodate multi-channel inputs for a fair comparison. \red{Methods and our corresponding modifications are introduced in Appendix \ref{app:method_selection}}.

\subsection{Our Method: AuthG-Live \& AuthG-Net}
\label{sec:our_method}
In this section, we \red{briefly introduce our liveness detection method, AuthG-Live, and our authentication model, AuthG-Net.}

\subsubsection{Feature Extraction}
\label{sec::our_method_feature_extraction}
\begin{redblock}
For feature extraction, log-Mel spectrograms were computed for both air-conduction (AC) and bone-conduction (BC) channels. BC signals were preprocessed to suppress low-frequency interference, while all signals were resampled to a unified sampling rate prior to transformation. The Mel spectrograms were generated using a fixed set of Mel filter banks within a predefined frequency range.

To characterize sound field (SF) properties, we extracted three complementary features: Energy Ratio over Time (ERT), Time Delay over Time (TDT), and Energy Distribution over Frequency (EDF). These features capture spatial sound energy attenuation, inter-channel propagation delays, and spectral energy distribution across microphones, following the design principles described in Section~\ref{sec::chap3_validation_acoustic}.

ERT quantifies relative energy differences across channels by computing band-limited spectral energy and normalizing it with respect to reference AC channel (Channel 7), followed by scaling and normalization to ensure numerical stability. TDT estimates relative time delays between each channel and the reference channel via cross-correlation. A confidence-aware weighting mechanism is applied to suppress unreliable delay estimates, and the resulting values are further normalized for stability. EDF characterizes the frequency-wise energy distribution for each channel by aggregating spectral representations over time. The distributions are normalized with respect to the reference channel to preserve relative spectral characteristics while reducing sensitivity to overall amplitude variations. Detailed implementation and parameter settings are provided in Appendix~\ref{app:feature_construction}.
\end{redblock}

\subsubsection{AuthG-Live: Attack Device Independent Voice Liveness Detection}
\begin{figure}
    \centering
    \includegraphics[width=0.8\linewidth]{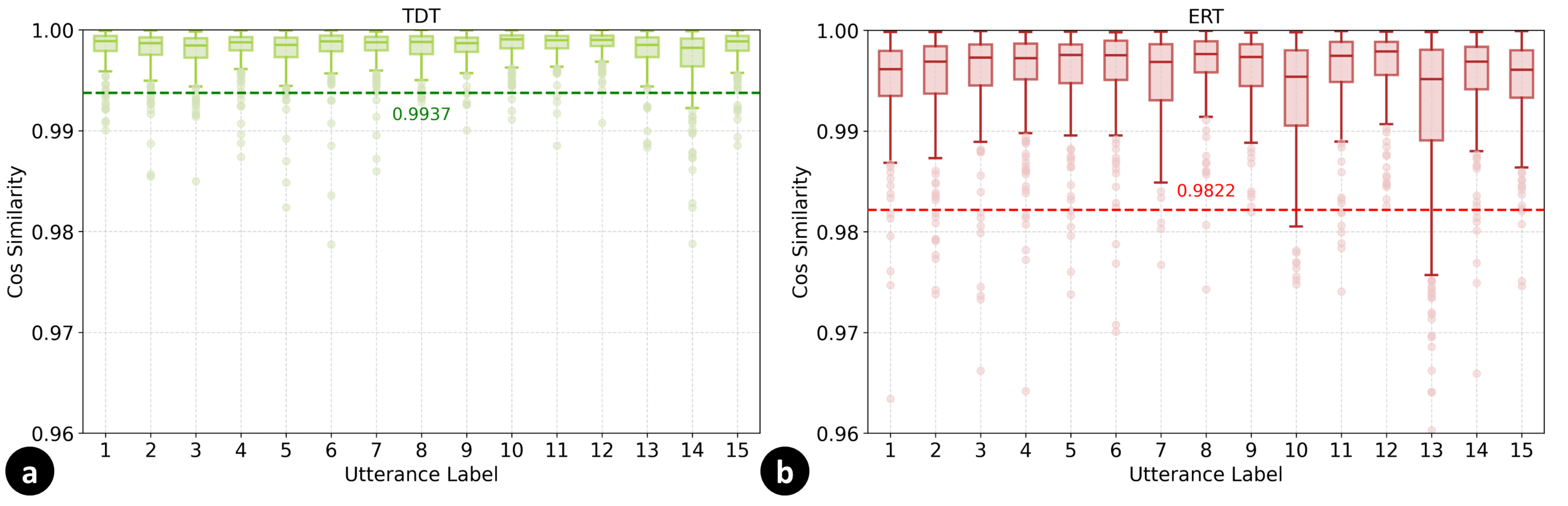}
    \caption{This figure shows the distribution of cosine similarities between the unified patterns and the corresponding aTDT and aERT across all genuine samples for the 15 utterances. Each box in the boxplot represents the interquartile range (25th to 75th percentile), with the line inside indicating the median. Scattered points denote outliers, and the dashed line marks the 3rd percentile of the genuine similarity distributions. In (a), the TDT cosine similarity between aTDT and the unified patterns defines a threshold of 0.9937, whereas in (b), the ERT yields a threshold of 0.9822.}
    \label{fig:TDT ERT CS}
\end{figure}

Most existing voice liveness detection methods learn discriminative patterns from attack samples observed during training~\cite{ahmed2020void, he2024fast}. As a result, their generalization performance strongly depends on the diversity of attack types seen during training. When confronted with unseen spoofing techniques or replay configurations, detection accuracy often degrades, leading to reduced robustness in real-world deployments.

\red{In contrast, AuthG-Live is a generalizable voice liveness detection framework that leverages physical sound propagation characteristics rather than attack-specific artifacts. It models the near-source spatiotemporal sound field by extracting Energy Ratio over Time (ERT) and Time Delay over Time (TDT) from multi-channel air-conducted signals, capturing intrinsic source–vocalization relationships that are difficult for replay or synthesized attacks to replicate. Notably, AuthG-Live is trained solely on genuine speech data, avoiding the need to enumerate diverse attack scenarios and enabling robustness to unseen spoofing methods. Furthermore, these features can be extracted in a streaming manner, supporting rapid attack detection (Section~\ref{sec::chap5_eval_live}).}

To this end, we incorporate TDT and ERT, as these features reflect the spatial origin of the received signals. Specifically, TDT and ERT are averaged along the time axis for each channel to form mean templates, denoted as aTDT and aERT, which summarize sound propagation delay and energy attenuation patterns. For liveness detection, the aTDT and aERT from genuine user samples are averaged to obtain unified aTDT and aERT patterns. These two unified patterns represents the typical acoustic propagation behavior of genuine speech. Given an input signal, its corresponding aTDT and aERT are computed and compared with the unified patterns using cosine similarity. A similarity score exceeding a predefined threshold indicates that the signal is likely emitted from a real human speaker.

To determine the decision threshold, we compute the cosine similarity between the unified patterns and the corresponding aTDT and aERT extracted from all genuine samples. The results exhibit consistently high similarity across utterances and subjects. Accordingly, the thresholds are set based on the 3rd percentile of the genuine similarity distributions, allowing 97\% of genuine samples to be accepted. This data-driven criterion ensures a high true acceptance rate for genuine speech while effectively rejecting most spoofing attacks. As illustrated in Fig.~\ref{fig:TDT ERT CS}, the resulting similarity thresholds are 
0.9937 for aTDT and 
0.9822 for aERT. An input signal is classified as live only if both similarity scores exceed their respective thresholds.

\subsubsection{AuthG-Net: Passphrase-Independent Voice Authentication}

\begin{figure}
\centering
\includegraphics[width=1.0\linewidth]{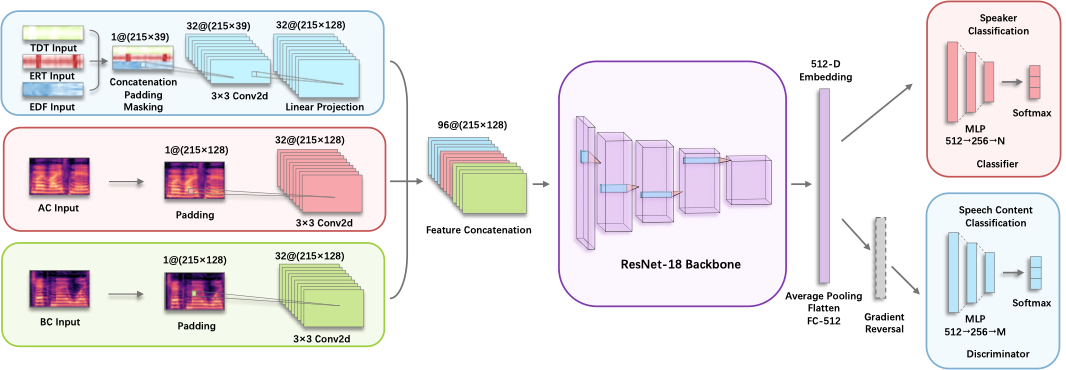}
\caption{The figure demonstrate the architecture of AuthG-Net. The blue/red/green blocks on the left denote the SF/AC/BC inputs, which are encoded by lightweight CNN heads and then stacked and fed into a ResNet-18 backbone (purple block) to produce a 512-D embedding. In the domain-adversarial learning (DAL) branch, the model is optimized with two objectives: speaker classification (red) and speech-content classification (blue). The speaker classifier is trained directly on the embedding, whereas the passphrase classifier is connected via a Gradient Reversal Layer (GRL), which reverses gradients during backpropagation to encourage content-invariant embeddings.
}
\label{fig:AuthG_model}
\end{figure}

\begin{redblock}
AuthG-Net is a lightweight multi-modal neural network designed for user authentication on smart glasses, as illustrated in Fig.~\ref{fig:AuthG_model}. It takes air-conducted (AC), bone-conducted (BC), and sound-field (SF) features as input, and learns a unified user representation through multi-modal fusion.

Specifically, each modality is first processed by a dedicated encoder to extract modality-specific features. These features are then fused and fed into a shared backbone network to produce a compact embedding representing the user. To enable generalization across different utterances, we adopt a domain-adversarial learning (DAL) framework, where the model is jointly optimized for speaker classification while suppressing speech-content information via an adversarial objective. This design encourages the learned embedding to capture user-specific characteristics that are invariant to the spoken content.

During training, the model is supervised with both subject and utterance labels to learn discriminative and utterance-independent representations. During inference, the extracted embedding is used as the user template, and authentication is performed by comparing embeddings using cosine similarity. Detailed network architecture, feature processing, and training configurations are provided in Appendix~\ref{app:authG-Net_implementation}.
\end{redblock}

\subsection{Results and Analysis on Benchmark Tasks}
\label{sec:benchmark_result}
In this section, we provide result and analysis on the benchmark tasks defined in section \ref{sec:benchmark_tasks_design}

\begin{figure}
    \centering
    \includegraphics[width=0.8\linewidth]{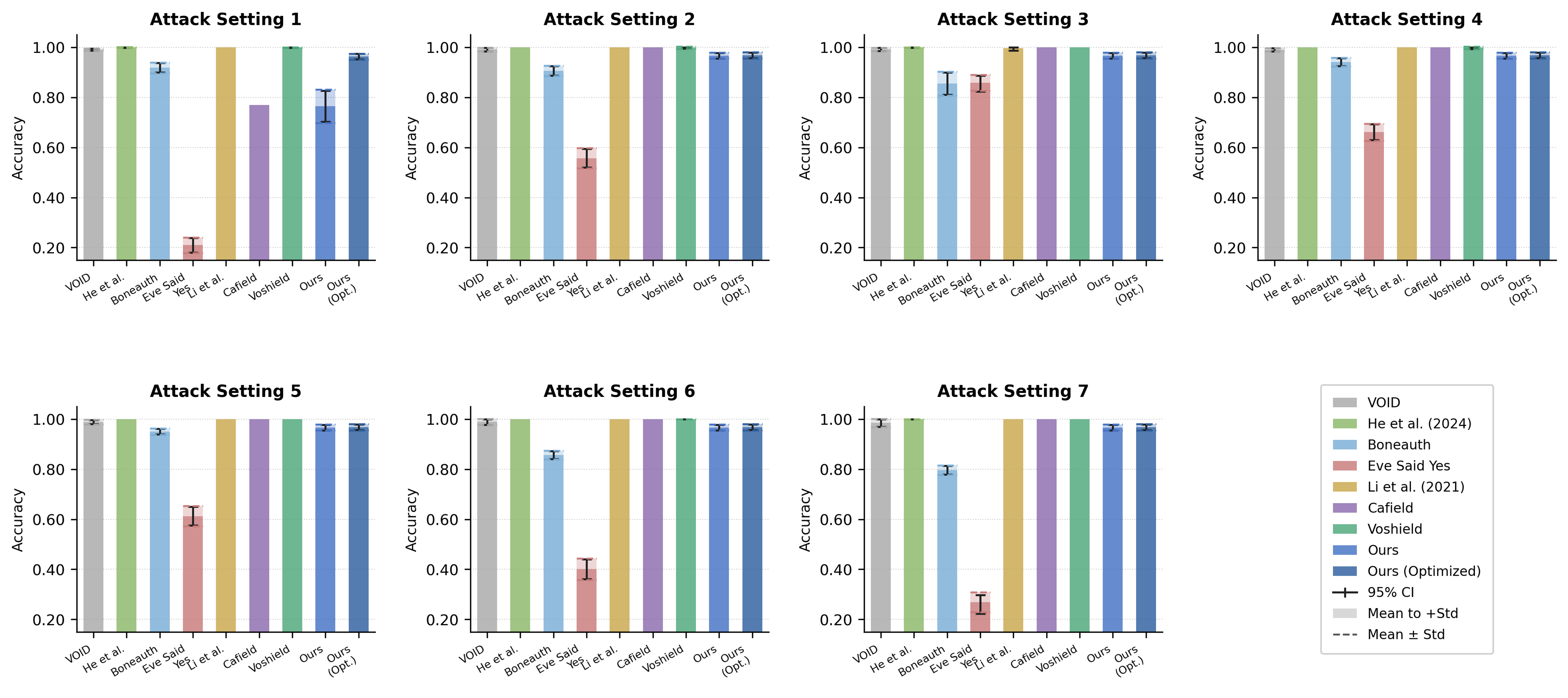}
    \caption{Result for benchmark task 1. \red{Note: (1) optimization for our method is described in section~\ref{sec:limited_channels}; (2) Since the other methods are trained and evaluated using known attack types, they may be prone to overfitting to specific attacks. In contrast, our method determines the decision threshold using only genuine samples and demonstrates superior performance under unseen attack scenarios, as discussed in Section~\ref{sec::benchmark_result_task2}.}}
    \label{fig:table7}
\end{figure}

\subsubsection{Task 1: Liveness Detection Accuracy.}
\label{sec::benchmark_result_task1}
We first evaluate all selected liveness detection methods on Task~1 using 7-fold cross-validation, with results summarized in Fig.~\ref{fig:table7}. The result for original baseline methods are also included for reference (Appendix~\ref{sec:performance_of_the_original_liveness}). Our method, leveraging all SF modalities, achieves high accuracy for Attack Setting~2 to 7, demonstrating strong robustness against environment-based replay attacks.

However, for Attack Setting~1 (wearer-based attack), the accuracy drops to 76\%, indicating that the attacker can partially mimic the spatial characteristics of a genuine user. As analyzed in Section~\ref{sec:limited_channels}, this degradation arises from channels that are insensitive to sound source position when relying on time delay and energy ratio cues. By selecting only spatially sensitive channels, the accuracy improves significantly to over 96\%.

Although several baselines report near-perfect accuracy, this performance likely stems from learning device-specific characteristics of the attacking loudspeaker rather than intrinsic liveness cues. Since each experiment uses the same attack type for both training and testing, models can exploit device-related features. As shown in later generalization experiments, such methods degrade substantially when evaluated on unseen attacks.

Finally, \textit{Eve Said Yes}, which relies on loose AC–BC similarity, fails to distinguish genuine and attack samples. This is because airborne signals can also induce correlated vibrations on BC sensors, and similar AC–BC patterns can be reproduced in simulated wearer attacks, making reliable discrimination difficult.

\begin{figure}
    \centering
    \includegraphics[width=0.8\linewidth]{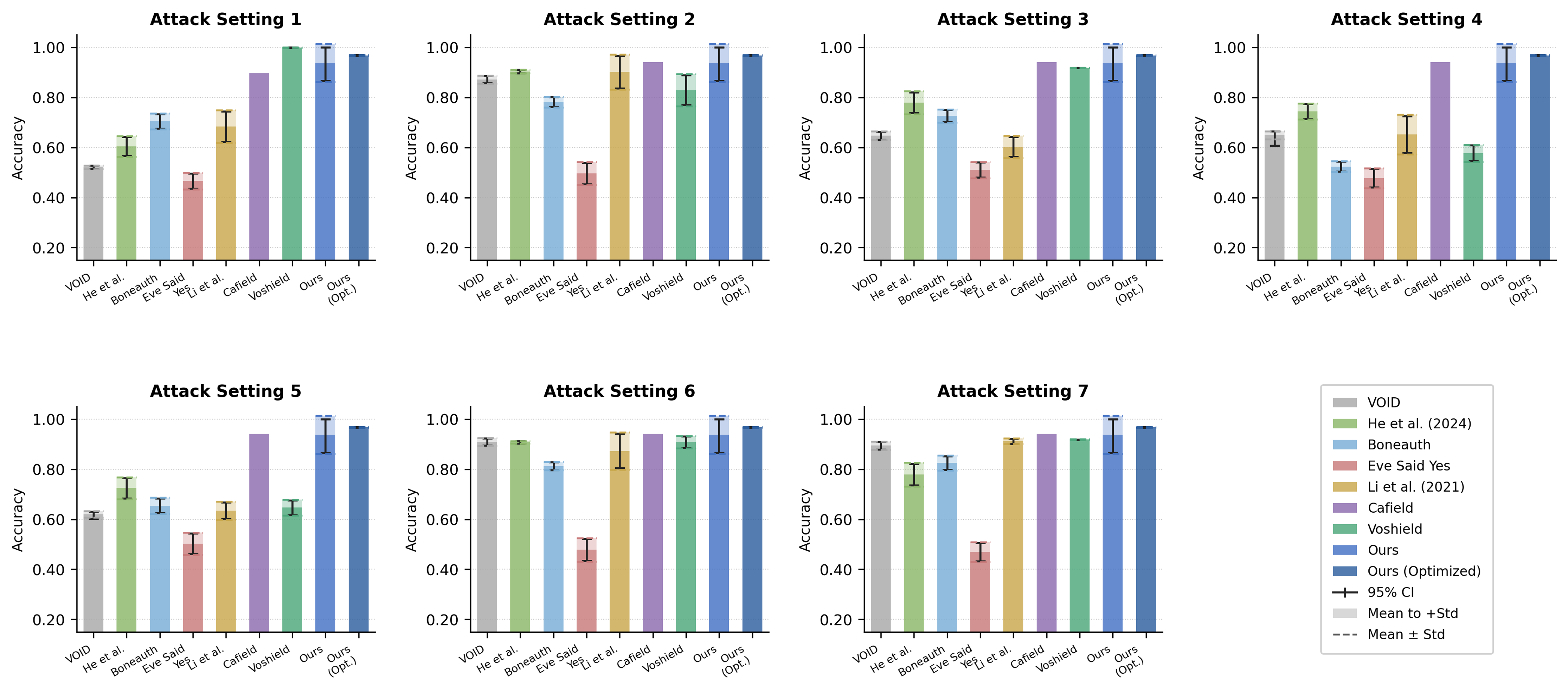}
    \caption{Result for benchmark task 2. \red{Note: (1) optimization for our method is described in section~\ref{sec:limited_channels}; (2) As our method do not require attack samples for training, we average the liveness detection accuracy on all attacks.}}
    \label{fig:table8}
\end{figure}

\subsubsection{Task 2: Generalization of Liveness Detection to Unseen Attacks.}
\label{sec::benchmark_result_task2}
We further evaluate generalizability through Task~2 by training on one attack setting and testing on others with a \red{7-fold cross-validation}. Results are shown in Fig.~\ref{fig:table8}. Noticeably, all baseline methods suffer performance degradation when facing unseen attacks, with those relying on attacker-device-specific features (e.g., VOID and He et al.) degrading the most.

By enforcing a strict AC–BC similarity metric, Boneauth retains partial attack detection capability, but still drops from around 90\% to 75\%, as its thresholds (determined by EER) vary across attack \red{settings} and become inconsistent. \textit{Eve Said Yes} completely fails on unseen attacks, yielding around 50\% accuracy. In contrast, SF-based methods (Li et al., Cafield, and Voshield) exhibit less degradation, suggesting that sound field features support cross-attack detection since they are independent of attack-device characteristics.

Among Li et al., Cafield, Voshield, and our method, ours (with optimization) achieves the highest and most stable accuracy across all seven attack types, while the others degrade under one or more attacks. Li et al., which uses original STFT spectrograms, inevitably captures device-specific features and shows the largest degradation (maximum over 30\%) compared to Cafield and Voshield. Although the latter two rely only on SF-related features, their CNN-based models still require both genuine and attack samples during training, limiting generalization. In contrast, our method requires no attack samples and relies on static similarity-based comparison, enabling robust performance on unseen attacks.

\begin{figure}
    \centering
    \includegraphics[width=0.6\linewidth]{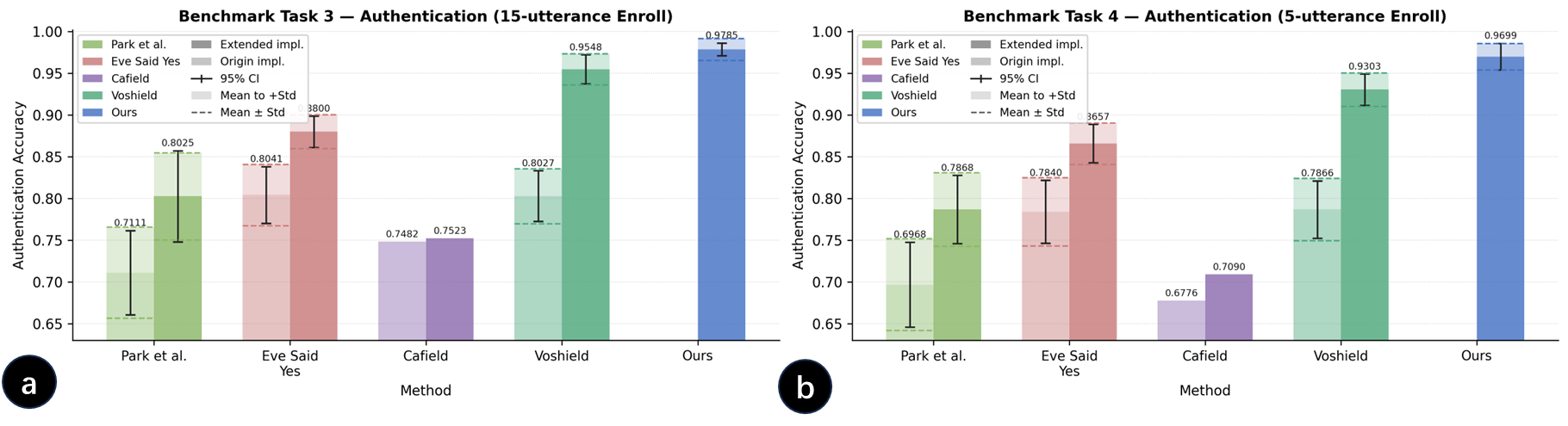}
    \caption{(a) Result for benchmark task 3. (b) Result for benchmark task 4.}
    \label{fig:table9}
\end{figure}

\subsubsection{Task 3: Authentication Accuracy}
\label{sec::benchmark_result_task3}
We evaluate authentication performance in Task~3 using 7-fold cross-validation, with results shown in Fig.~\ref{fig:table9}a. Our method achieves the best performance among all compared approaches, demonstrating that the proposed feature extraction strategy and AuthG-Net can effectively capture user-related characteristics.

Park et al. employ an LSTM with randomized embedding generation, but their method is designed for single-user scenarios and fails to learn discriminative embeddings across multiple users, resulting in low accuracy (80.25\%). Although \textit{Eve Said Yes} supports embedding extraction, it struggles to effectively fuse multi-modal features, yielding only limited improvement (from \red{80.41\%} to \red{88.00\%}). Cafield utilizes fieldprint SF features but performs poorly under both original and extended implementations, likely due to the limited capacity of its GMM-based classifier.
In contrast, Voshield, based on SFD features, achieves performance close to ours, improving by approximately \red{13\%} over its original implementation, which indicates the effectiveness of CNN-based architectures in extracting discriminative acoustic representations.

\subsubsection{Task 4: Cross-Utterance Authentication Performance}
We evaluate cross-utterance authentication performance in Task~4 using 7-fold cross-validation. As shown in Fig.~\ref{fig:table9} b, our method achieves the best performance, with only ~1\% degradation compared to Task~3, indicating its ability to learn utterance-independent user features, largely due to DAL.

In contrast, the other methods exhibit varying degradation, with Cafield and Voshield being most affected (5\% and 3\%, respectively). This is likely because their sound field features (fieldprint and SFD) are closely tied to utterance content, as different phonemes produce distinct sound field characteristics. By separately modeling sound field features in both frequency (EDF) and time domains (ERT and TDT), our method achieves stronger generalization across utterances.
For Park et al. and \textit{Eve Said Yes}, although their overall authentication performance is relatively limited, the degradation under cross-utterance settings is smaller (around 1\%–2\%), suggesting partial robustness in extracting utterance-independent features.

\subsection{Conclusion}
\label{sec:authentication_conclusion}
In conclusion, our method achieves state-of-the-art performance in both voice liveness detection and authentication. It demonstrates strong generalization capability for liveness detection against unseen attacks, as well as robust performance in cross-utterance authentication scenarios.
\section{Practical Evaluations}
\label{:Evaluation}
Since our method is developed using high-quality equipment for theoretical analysis, it is necessary to validate its effectiveness under practical constraints. To this end, we conduct a series of ablation studies. Specifically, we evaluate the impact of reduced sampling rates compared to the original 96,kHz setting (Section~\ref{sec:limited_sampling_rate}), limited acoustic modalities instead of jointly using AC, BC, and SF (Section~\ref{sec:limited_acoustic_modality}), and a reduced number of SF channels given the impracticality of deploying 14 AC microphones on smart glasses (Section~\ref{sec:limited_channels}).
We further simulate microphone layouts of off-the-shelf products, including Rokid~\cite{RokidGlasses2025} and Meta Ray-Ban~\cite{MetaRayban2025}, to evaluate both liveness detection and authentication under realistic settings (Section~\ref{sec:real_world_products}). \red{Finally, we conduct a simulated live study to assess liveness detection and authentication under two levels of system simplification, focusing on latency, energy consumption, and performance (Section~\ref{sec::chap5_eval_live}).}

\subsection{Limited Sampling Rate}
\label{sec:limited_sampling_rate}
\begin{figure}
    \centering
    \includegraphics[width=0.8\linewidth]{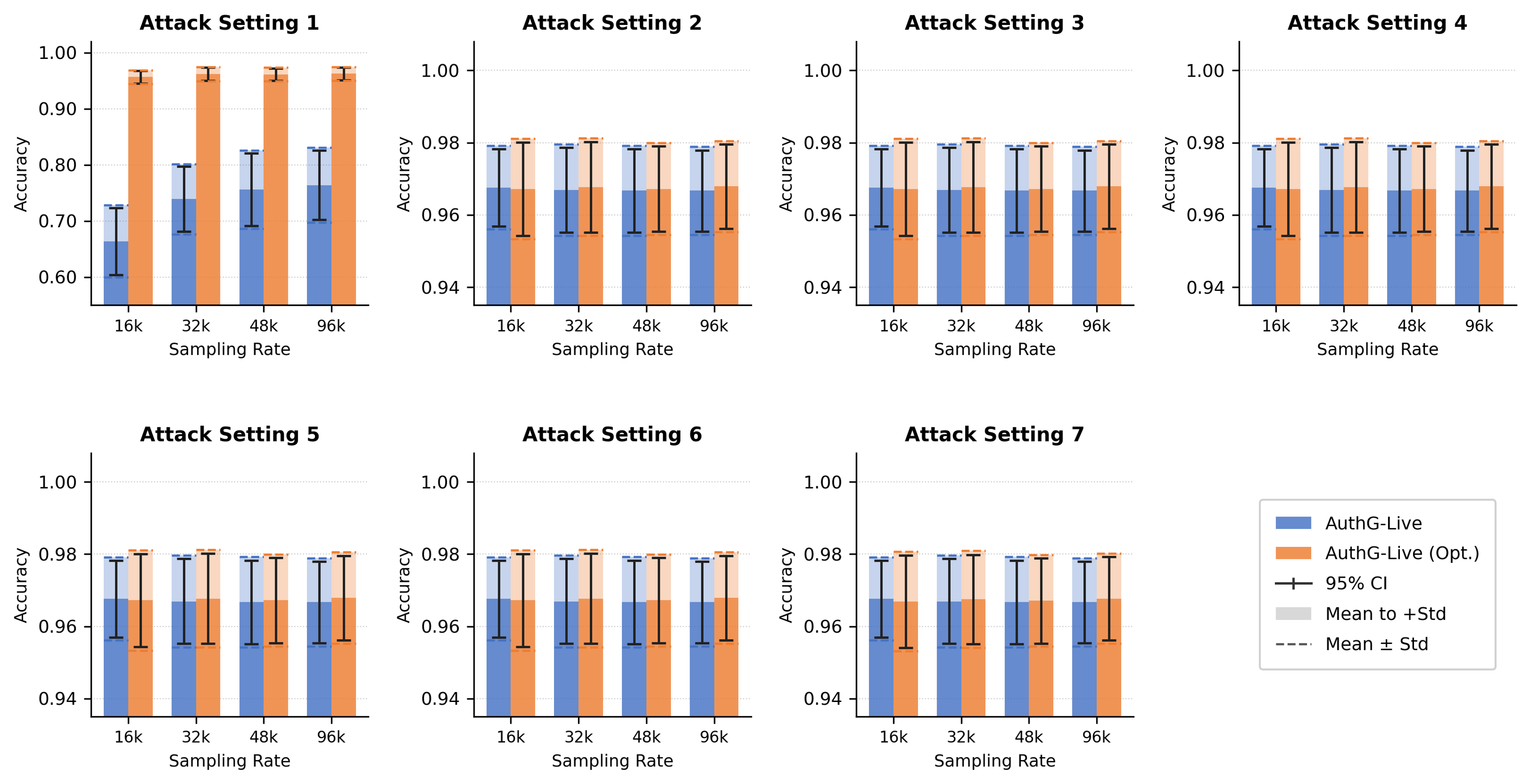}
    \caption{Liveness detection accuracy under different sampling rates. Note: Optimization is described in section~\ref{sec:limited_channels}}
    \label{fig:table10}
\end{figure}

\begin{figure}
    \centering
    \includegraphics[width=0.6\linewidth]{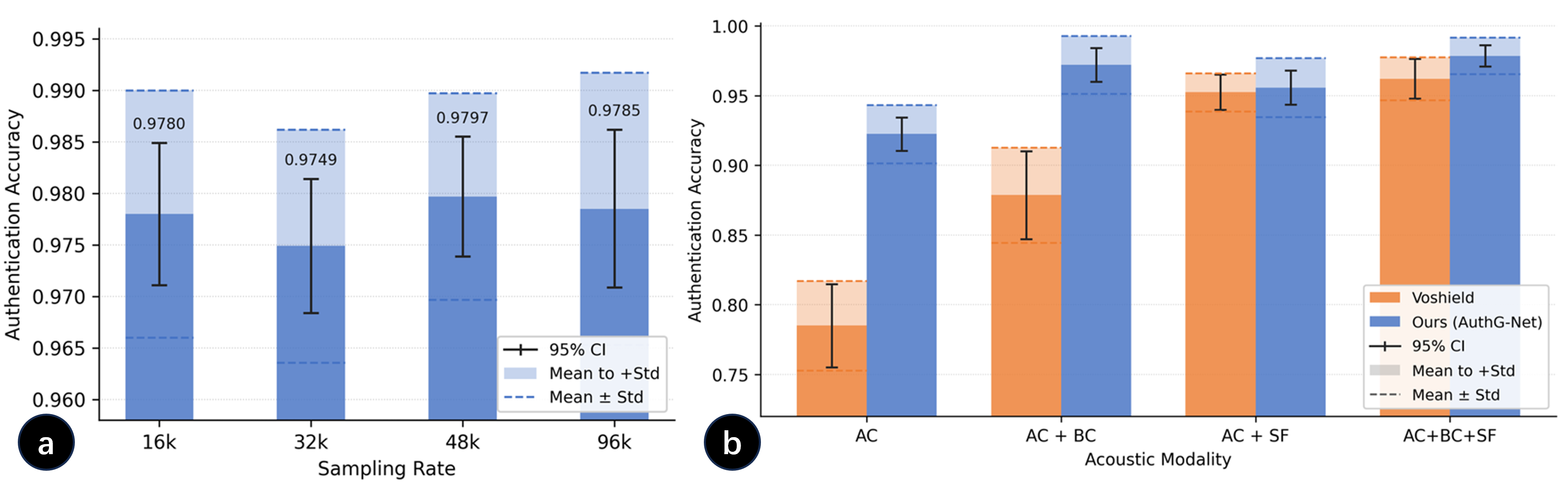}
    \caption{(a) Authentication accuracy under different sampling rates. (b) Authentication accuracy under different input modalities.}
    \label{fig:table11-12}
\end{figure}

Our prototype records audio at 96~kHz, whereas commercial smart glasses typically operate at around 16~kHz. To evaluate the impact of sampling rate, we test four settings: 16~kHz, 32~kHz, 48~kHz, and 96~kHz. Lower-rate recordings are simulated by applying FIR-based low-pass filtering followed by uniform downsampling on each channel.
While reduced sampling rates have minimal effect on Mel-spectrogram computation, they can degrade sound field modeling due to reduced temporal resolution. To mitigate this, the downsampled signals are linearly interpolated to recover a denser temporal representation while preserving bandwidth constraints. The processed signals are then used for feature extraction and evaluation.
We follow Benchmark Task~1 (liveness detection) and Task~3 (authentication) in Section~\ref{sec:benchmark_tasks_design}, \red{using 7-fold cross-validation repeated twice}. All AC, BC, and SF modalities are included. Results are shown in Fig.~\ref{fig:table10} (liveness detection) and Fig.~\ref{fig:table11-12}a (authentication).

\subsubsection{Results and Discussion}
As shown in Fig.~\ref{fig:table10}, the original AuthG-Live with all SF channels achieves consistently high liveness detection accuracy on attack settings 2–7 (all above 96\%), with no noticeable degradation as the sampling rate decreases. However, it performs poorly on attack setting 1, where accuracy drops from 76\% to 66\% as the sampling rate decreases from 96~kHz to 16~kHz.
Similar with analysis in Section~\ref{sec::benchmark_result_task1}, this issue can be addressed by the channel optimization introduced in Section~\ref{sec:limited_channels}, which improves robustness by focusing on position-sensitive SF channels. With this optimization, AuthG-Live achieves consistently high accuracy (above 95\%) across all seven attack settings under different sampling rates;

For authentication, Fig.~\ref{fig:table11-12}a shows that AuthG-Net maintains strong performance despite reduced sampling rates, achieving accuracy above 97.93\%.

\subsection{Limited Acoustic Modalities}
\label{sec:limited_acoustic_modality}
Although our prototype supports AC, BC, and SF modalities, many commercial smart glasses lack BC sensors or sufficient microphones for SF capture. It is therefore important to evaluate whether our method remains effective under limited modality configurations.

We consider four representative combinations: AC only, AC+BC, AC+SF, and AC+BC+SF. Since AuthG-Live relies solely on SF and cannot operate without SF inputs, we evaluate only authentication performance using AuthG-Net, following Benchmark Task~3 (Section~\ref{sec::benchmark_task3}) \red{with 7-fold cross-validation repeated twice}. The sampling rate is fixed at 96~kHz. To support different modality inputs, we implement three additional variants of AuthG-Net by removing the corresponding modality encoders before feature fusion, leveraging its modular design. For example, the AC+SF variant retains only AC and SF encoders, while the AC-only variant reduces to a residual CNN with DAL.

We also implement Voshield for comparison, as it achieves the second-best performance in Section~\ref{sec::benchmark_result_task3}. Similarly, three variants of Voshield are constructed to support reduced modality inputs following the same modification principle (Section~\ref{sec:method_selection}).

\subsubsection{Results and Discussion}
As shown in Fig.~\ref{fig:table11-12}b, reducing available acoustic modalities degrades authentication performance for both our method and the baseline, confirming that BC and SF provide effective complementary information to AC.
Our method consistently outperforms the baseline when combining AC with either SF or BC, demonstrating that AuthG-Net can effectively extract and fuse user-related features across modalities. Notably, even with AC-only input, our method achieves \red{91.04\%} accuracy, still surpassing the baseline, indicating strong robustness under limited modality conditions.

\subsection{Limited Input Channels and Optimization on Channel Selection}
\label{sec:limited_channels}
After validating generalization across modality combinations, we further evaluate performance under limited input-channel settings, motivated by the practical constraints of microphone availability in smart glasses (e.g., power, weight, and comfort). We investigate whether reduced microphone configurations can achieve competitive or improved performance in both liveness detection and authentication. Specifically, we examine the number and placement of air-conduction (AC) microphones, as well as the impact of incorporating bone-conduction (BC) microphones.

We consider three representative AC configurations reflecting real devices: 4-channel (Rokid~\cite{RokidGlasses2025}), 5-channel (Ray-Ban Meta~\cite{MetaRayban2025}), and a feasible 7-channel setup based on our prototype. For each configuration, we optimize microphone placement for liveness detection using AuthG-Live, focusing on symmetric layouts commonly used in practice while also exploring asymmetric alternatives. During optimization, Channel~7 is fixed as the AC input, and all combinations of the remaining AC channels are evaluated following Benchmark Task~1 (Section~\ref{sec:benchmark_tasks_design}). The best-performing symmetric and asymmetric layouts are identified for each configuration.

Using these optimized layouts, we then evaluate authentication performance with and without BC inputs following Benchmark Task~3 (Section~\ref{sec:benchmark_tasks_design}) \red{with 7-fold cross-validation repeated twice}. To support varying SF channel numbers, we adjust the input dimension of the SF encoder projection layer in AuthG-Net, while optional BC inputs are incorporated as described in Section~\ref{sec:limited_acoustic_modality}. The sampling rate is fixed at 96~kHz.

\begin{figure}
    \centering
    \includegraphics[width=0.8\linewidth]{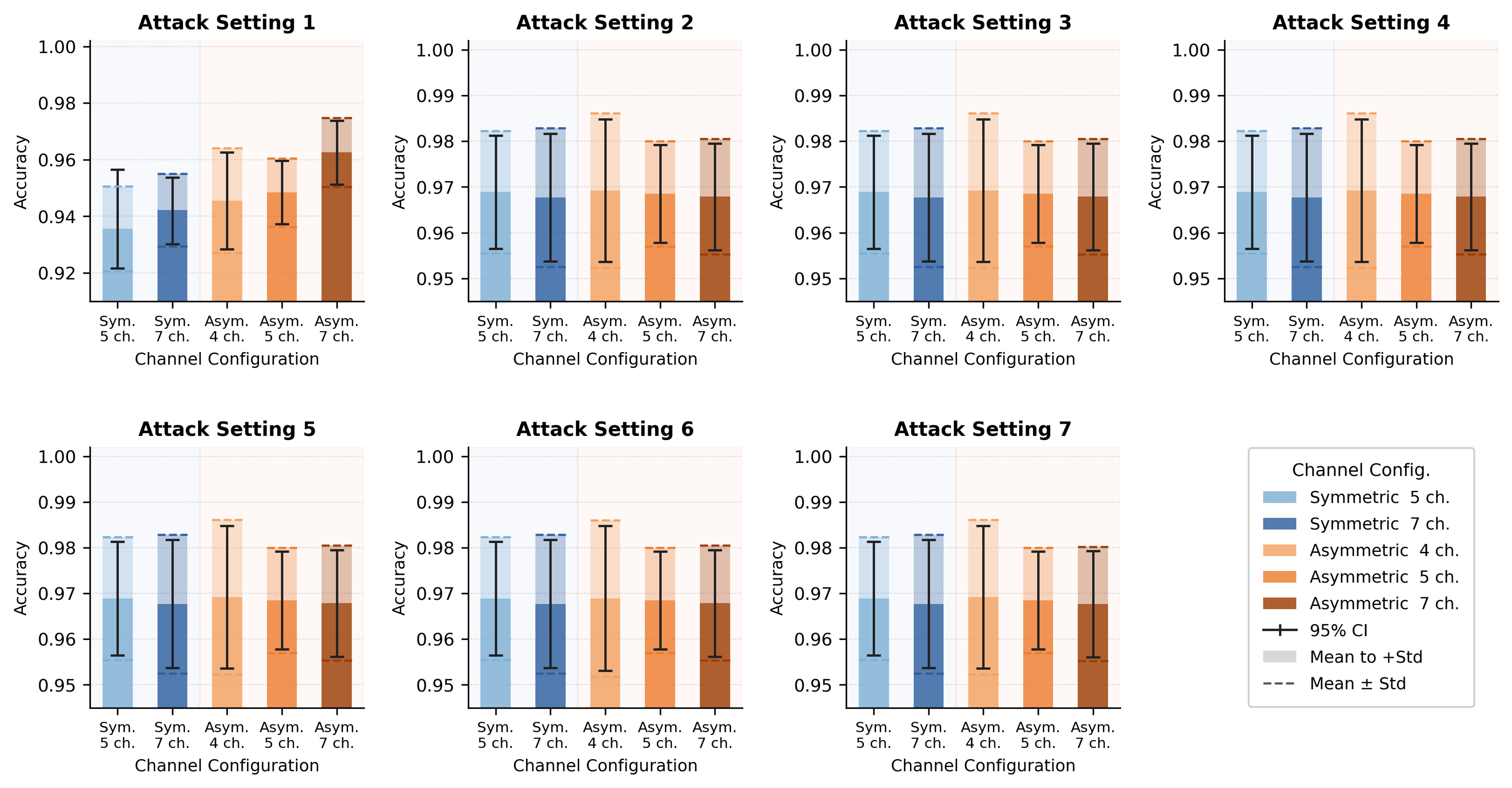}
    \caption{Liveness detection accuracy under different channel inputs.}
    \label{fig:table13}
\end{figure}

\begin{figure}
    \centering
    \includegraphics[width=0.6\linewidth]{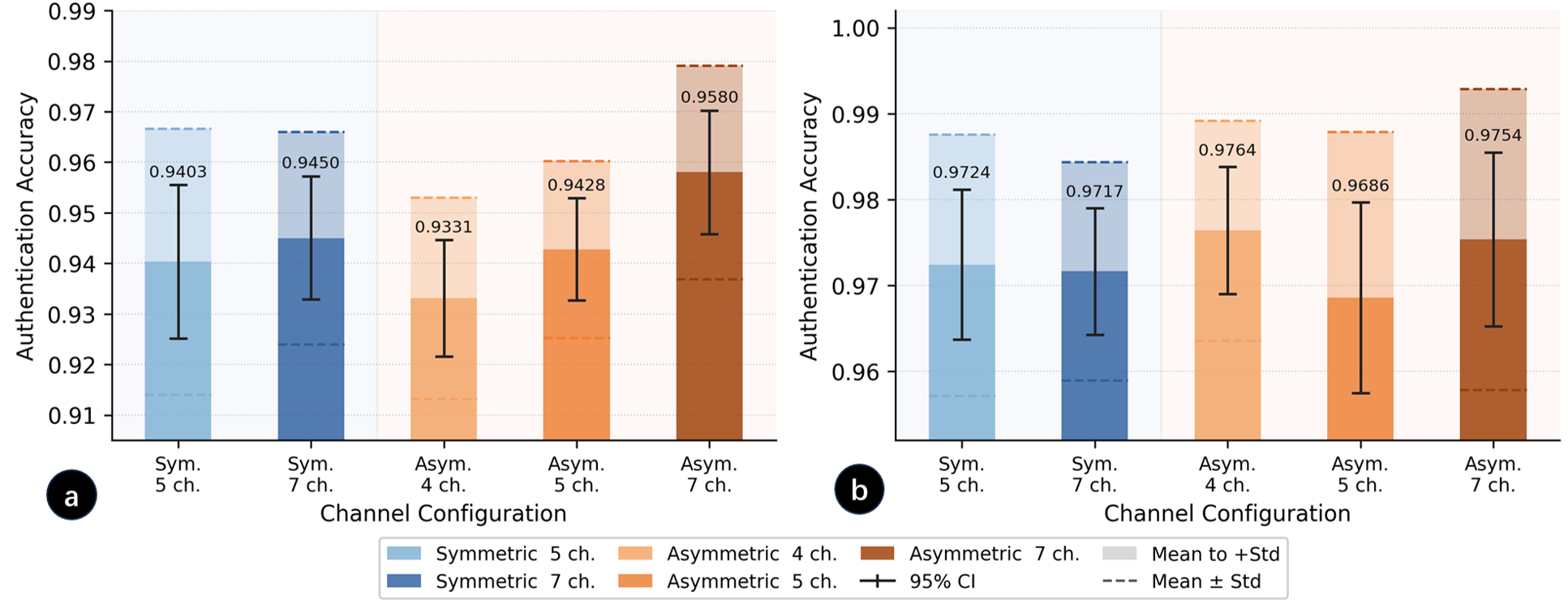}
    \caption{Authentication accuracy under different channel inputs (1) without and (2) with BC modality.}
    \label{fig:table14}
\end{figure}

\subsubsection{Results and Discussion}
\begin{figure}
    \centering
    \includegraphics[width=1.0\linewidth]{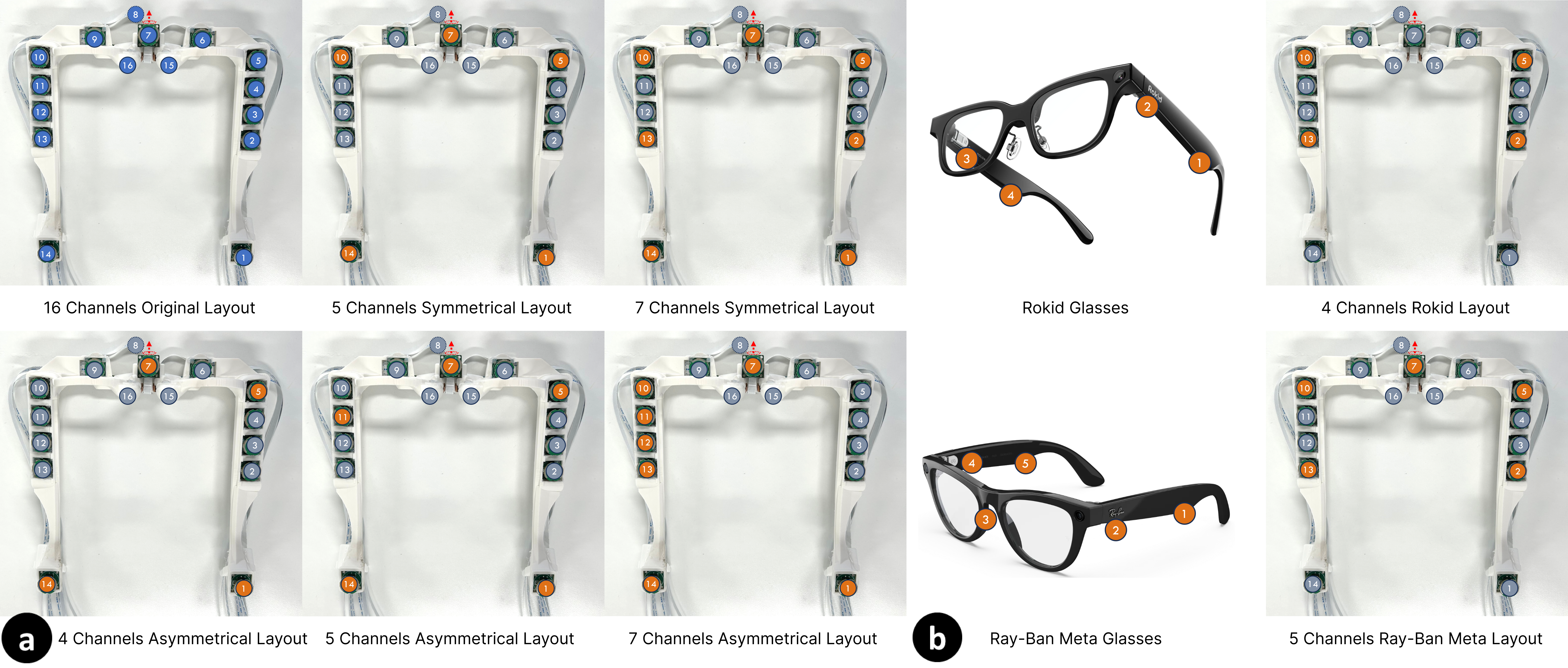}
    \caption{Liveness detection and authentication accuracy were evaluated under limited input-channel settings. (a) Top row shows the original microphone layout of the prototype and the selected microphones for the 5- and 7-channel symmetric configurations. Bottom row shows the selected microphones for the 4-, 5-, and 7-channel asymmetric configurations. Selected microphones are highlighted in orange. (b) demonstrates the microphone layouts of Rokid Glasses and Ray-Ban Meta Glasses on the left, and the microphones used in our experiments to simulate their recording characteristics on the right, with selected microphones highlighted in orange.}
    \label{fig:commercial_products}
\end{figure}

The optimized channel configurations are illustrated in Fig.~\ref{fig:commercial_products}. Across all setups, air-conduction microphones behind the ear are consistently selected, indicating higher sensitivity to time-delay and energy-ratio variations due to their distance from the mouth.

Fig.~\ref{fig:table13} shows the liveness detection accuracy of all configurations. For all setups, accuracy on attack settings 2 to 7 exceeds 96\%, demonstrating robust detection of environmental replay attacks. For attack setting 1, only the 7-channel asymmetric configuration achieves accuracy above 96\%, while others remain around 94\%, suggesting that both sufficient channel count and optimized placement are necessary to detect wearer-based attacks. Moreover, for the same number of channels, asymmetric layouts consistently outperform symmetric ones (e.g., 4-channel asymmetric vs. 5-channel symmetric), implying that asymmetric configurations capture richer spatial information given the approximate symmetry of human sound fields. These findings provide practical guidance for microphone deployment in smart glasses.

Fig.~\ref{fig:table14} demonstrates authentication accuracy with and without the BC modality. SF features from additional AC channels provide complementary information, with more channels generally improving performance. However, BC features offer a stronger complementary effect: when BC is included, accuracy remains stable across different SF configurations at around 97\%, consistently higher than configurations without BC. This suggests that incorporating BC microphones is more beneficial than relying solely on AC channels, offering further design insights for future smart glasses.

\subsection{Evaluation with Real-world Products' Configuration}
\label{sec:real_world_products}

\subsubsection{Rokid Glass}
Rokid Glass employs four symmetrically placed air-conduction microphones at the front and middle positions on both temples (Fig.~\ref{fig:commercial_products} b), operating at 16~kHz. To emulate this setup, we select Channels~2, 5, 10, and 13 in our prototype to match the spatial layout, with Channel~5 as the reference (substituted AC modality). Corresponding AC and SF features are recomputed following the modifications in Sections~\ref{sec:limited_sampling_rate} and~\ref{sec:limited_channels}. We evaluate liveness detection and authentication using Benchmark Task~1 and Task~3 (Section~\ref{sec:benchmark_tasks_design}), \red{with 7-fold cross-validation repeated twice}.

\subsubsection{Results and Discussion}
\begin{figure}
    \centering
    \includegraphics[width=0.8\linewidth]{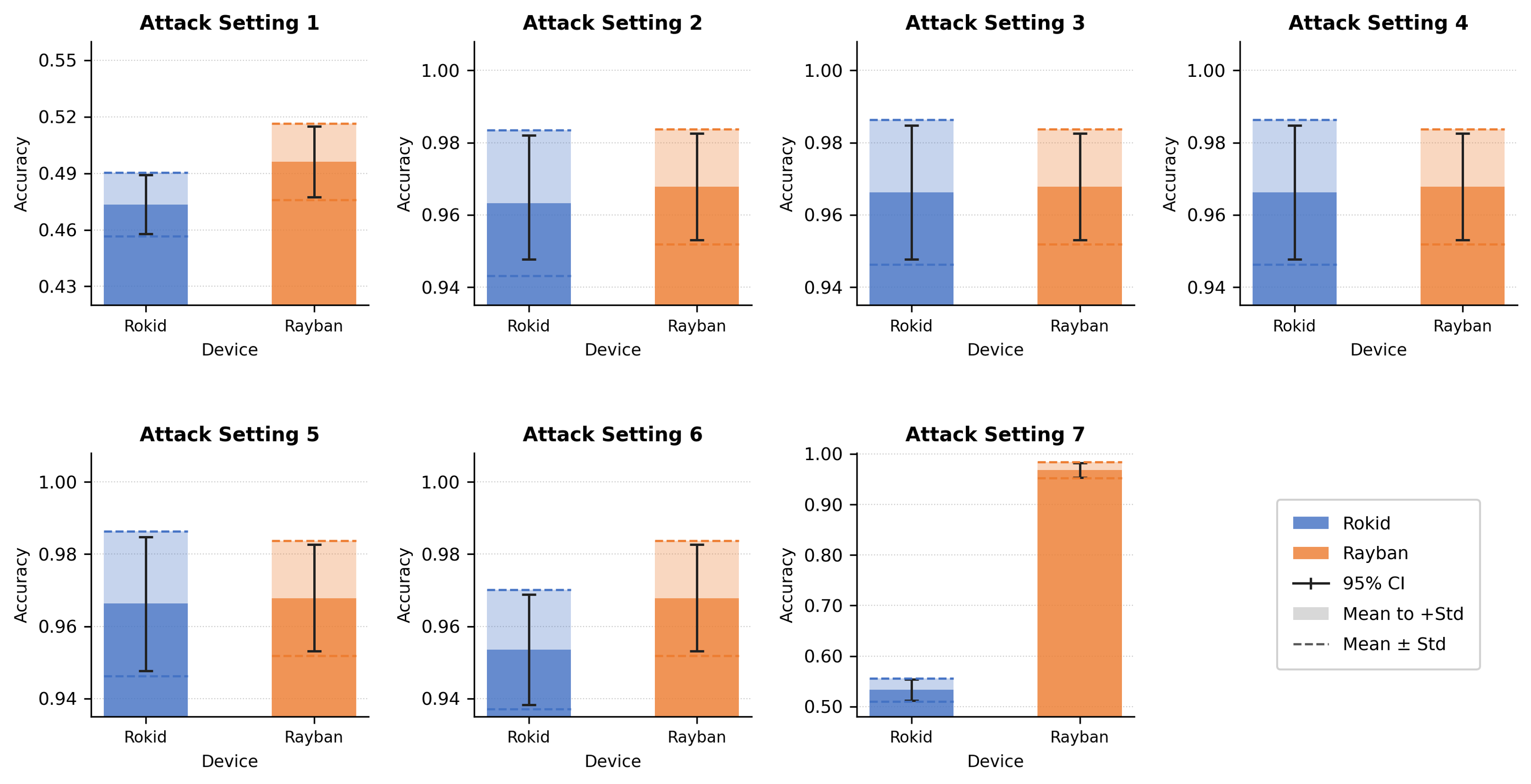}
    \caption{Liveness detection accuracy for Rokid and Ray-ban Meta configurations.}
    \label{fig:table15}
\end{figure}

Under the Rokid configuration, liveness detection accuracy (Fig.~\ref{fig:table15}) remains high for attack settings 2–6 (above 96\%), consistent with Section~\ref{sec:limited_channels}. However, performance drops significantly for attack settings 1 and 7 (i.e., closest or wearer-based attacks), with accuracy around 50\%. \red{This is likely due to the symmetric microphone layout introducing spatial redundancy and the compact distribution limiting sound field coverage: for sound sources close to the 4-channel array and near the symmetry plane, microphones on both sides receive highly similar signals, causing the aTDT and aERT features to collapse into a single three-point spike (upward or downward) that fails to capture the full voice propagation pattern; adopting asymmetric layouts, as illustrated in Section~\ref{sec:limited_channels}, can mitigate this issue.}

\begin{figure}
    \centering
    \includegraphics[width=0.4\linewidth]{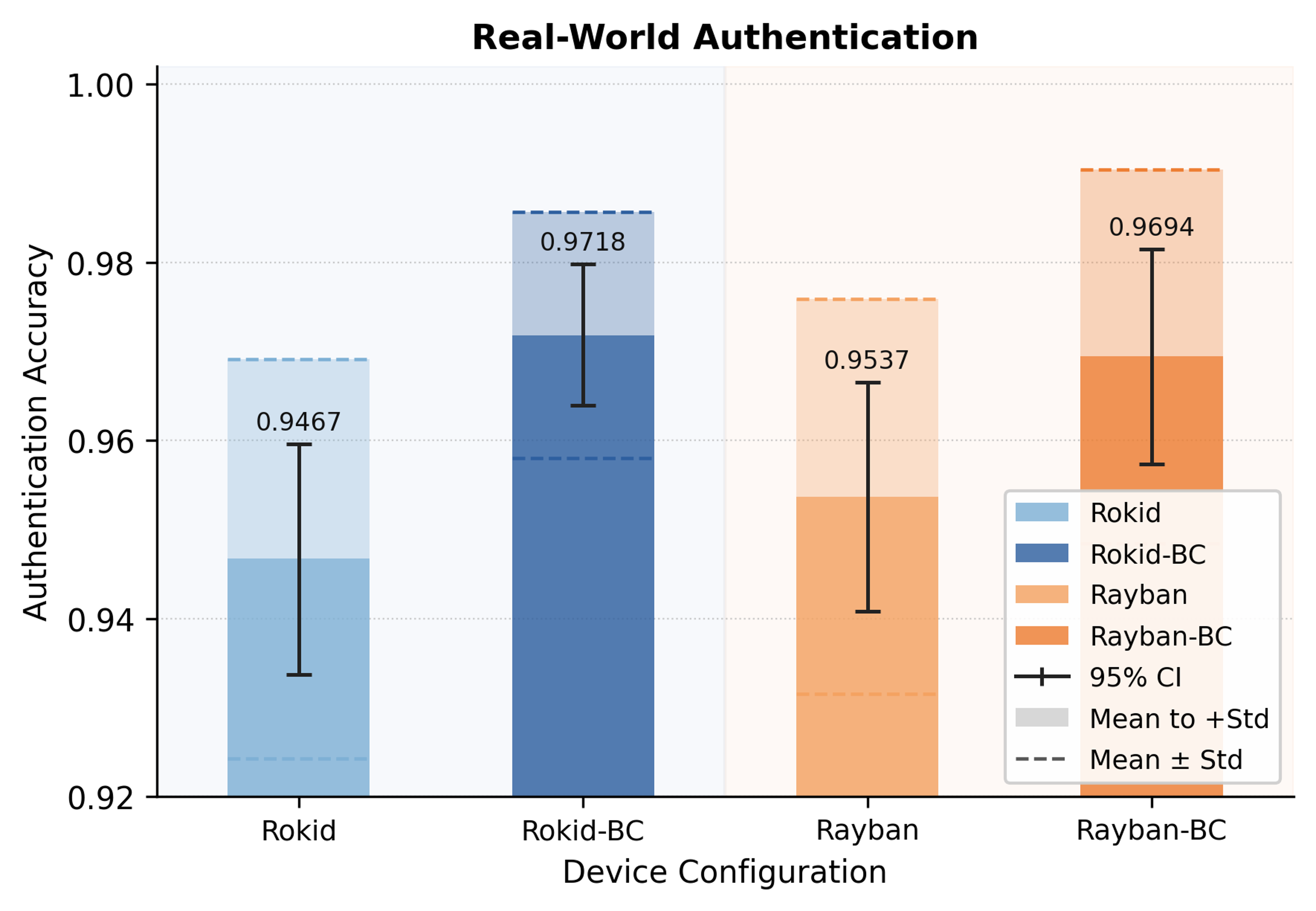}
    \caption{Authentication accuracy for Rokid and Ray-ban Meta configurations.}
    \label{fig:table16}
\end{figure}

For authentication (Fig.~\ref{fig:table16}), AuthG-Net achieves performance consistent with Section~\ref{sec:limited_channels}, reaching 97\% with BC and 94\% without BC. This demonstrates robustness to different microphone layouts, even when the AC channel is off-center, and confirms that AuthG-Net maintains high accuracy under reduced sampling rates (16~kHz vs. 96~kHz).

\subsubsection{Ray-Ban Meta}
Ray-Ban Meta integrates five air-conduction microphones at 16~kHz (Fig.~\ref{fig:commercial_products} b), including four similar to Rokid and one at the right nose pad. We emulate this setup using Channels~2, 5, 10, 13, and~7, with Channel~7 as the AC reference for SF feature extraction. Evaluation procedures and model modifications follow those of the Rokid configuration.

\subsubsection{Results and Discussion}
Under the Ray-Ban Meta configuration, liveness detection accuracy (Fig.~\ref{fig:table15}) remains high for attack settings 2–6 (above 95\%), consistent with Rokid. Unlike Rokid (\~50\%), performance on attack setting 7 is also high and comparable to Section~\ref{sec:limited_channels}. \red{This is because, with five channels, the aTDT and aERT features form a four-point trapezoidal pattern, which can partially capture the user’s voice propagation pattern.
However, performance on attack setting 1 remains around 50\%, indicating that when the sound source is near the mouth and exhibits head-like geometry, the symmetric 5-channel layout still fails to distinguish genuine speech from wearer-based attacks. Similar to Rokid, asymmetric microphone layouts may alleviate this limitation. Notably, the nose-pad microphone in Ray-Ban Meta introduces inherent asymmetry, which may enhance sound field feature capture; this effect warrants further investigation.}

For authentication (Fig.~\ref{fig:table16}), the simulated Ray-Ban Meta configuration at 16~kHz achieves performance comparable to both Rokid and Section~\ref{sec:limited_channels}, further confirming that AuthG-Net remains effective under limited sampling rates.

\subsection{Live Evaluation}
\label{sec::chap5_eval_live}
\begin{figure}
    \centering
    \includegraphics[width=0.4\linewidth]{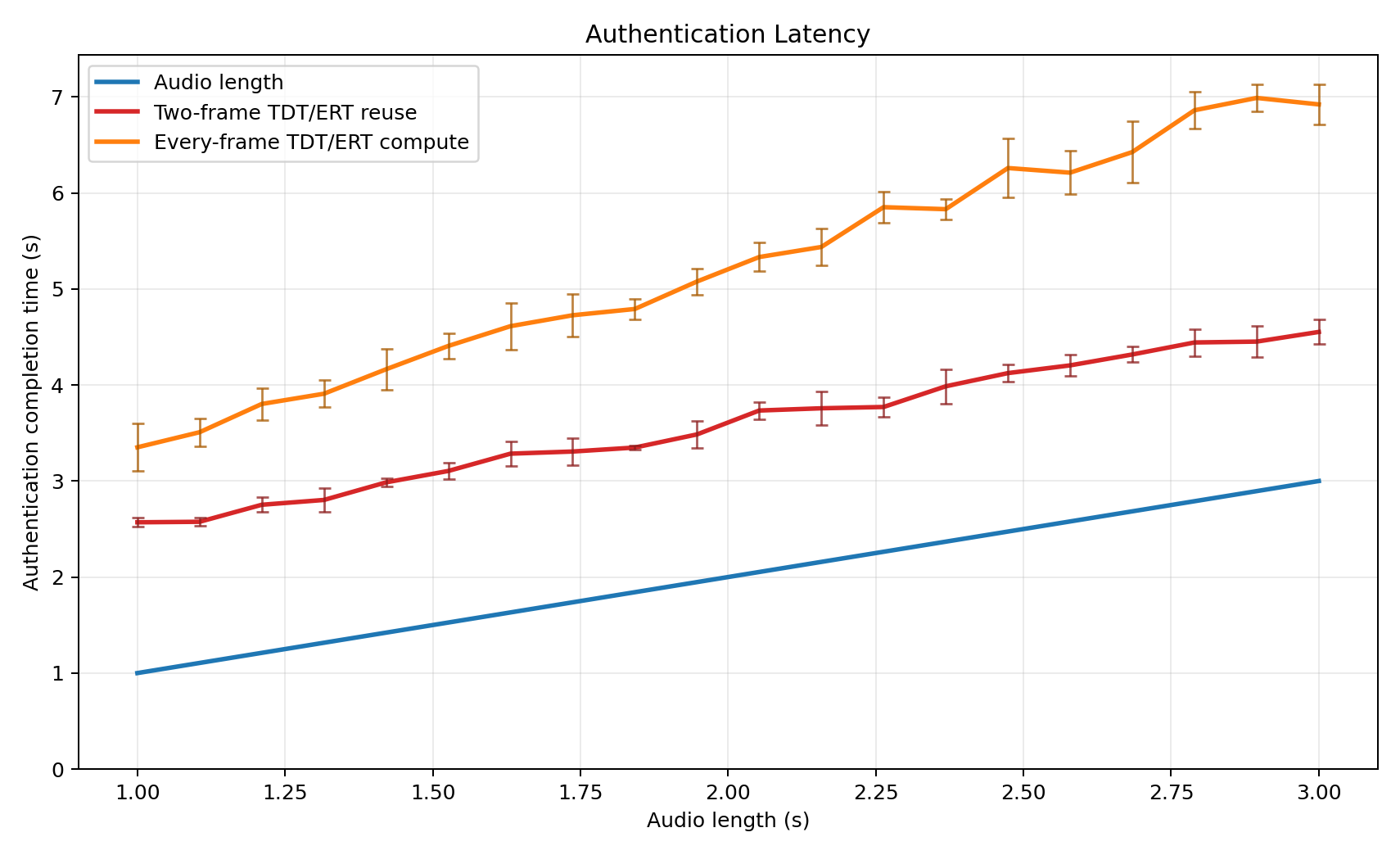}
    \caption{Live version latency under two levels of simplification}
    \label{fig:eval_live_latency}
\end{figure}

\begin{redblock}
\subsubsection{Implementation and Experiment Procedure}
We evaluate the feasibility of \textit{AuthG-Live} and \textit{AuthG-Net} under live deployment via a simulated setup. Since our acquisition board only supports synchronized recording, multi-channel audio is streamed and processed on a Raspberry Pi\cite{RaspBarryPi} to emulate a resource-constrained edge device. The system operates in a streaming manner according to the sampling rate, mimicking real-time processing.

\subsubsection{Live Simplification}
Preliminary results show that latency is dominated by TDT and ERT, which require frame-wise cross-correlation and energy computation. To reduce overhead, we design two simplification levels: \textbf{(a) Level 1:} compute TDT/ERT at 24~kHz; \textbf{(b) Level 2:} use odd frames with padding to reconstruct full samples, further reducing computation while preserving time-delay accuracy. All other features are processed with 96kHz.
We adopt a multi-threaded pipeline, where AC, BC, TDT, ERT, and EDF features are processed in parallel streams, followed by a unified inference thread for liveness detection and authentication.

\subsubsection{System Performance on Latency and Energy}
Latency is evaluated under simulated live conditions with 5 repeats for each length of input (Fig.~\ref{fig:eval_live_latency}). For Level~1, processing latency remains below 4~s for inputs shorter than 3~s, but grows with input length. In contrast, Level~2 achieves a nearly constant latency of around 1.5~s, independent of input duration, enabling near real-time processing.
On the other hand, energy overhead is moderate. The idle (no audio streaming) power consumption is 2.56~W. For each 3-second audio segment, feature processing increases power consumption by 2.32~W, while inference increases it by 4.87~W; averaged over the full authentication pipeline, the increase is 2.72~W. This additional cost is acceptable relative to the system's idle power baseline. For both Level~1 and Level~2, the memory footprint remains below 267~MB.

\begin{figure}
    \centering
    \includegraphics[width=0.8\linewidth]{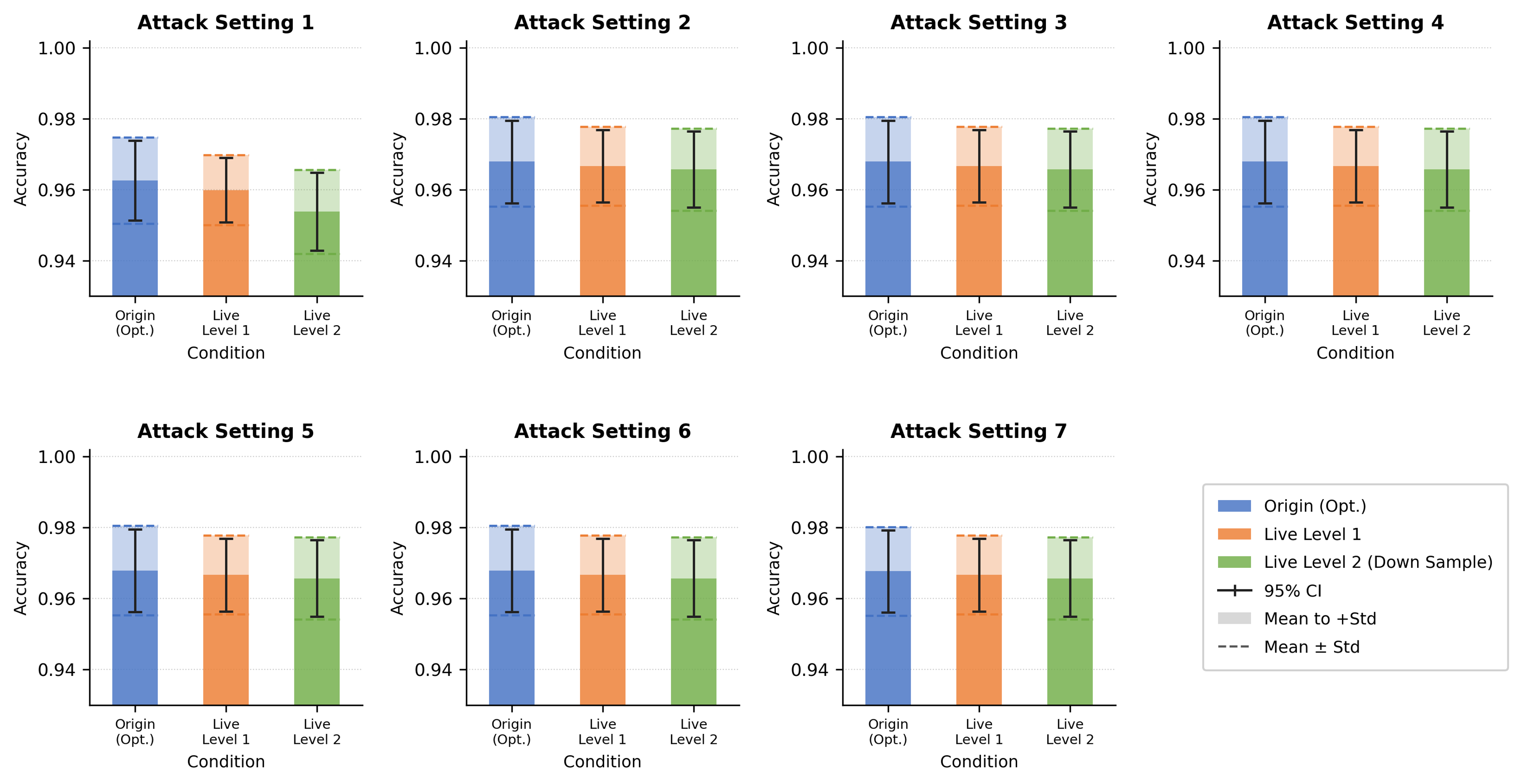}
    \caption{Liveness detection accuracy under two levels of simplification}
    \label{fig:table17}
\end{figure}

\begin{figure}
    \centering
    \includegraphics[width=0.5\linewidth]{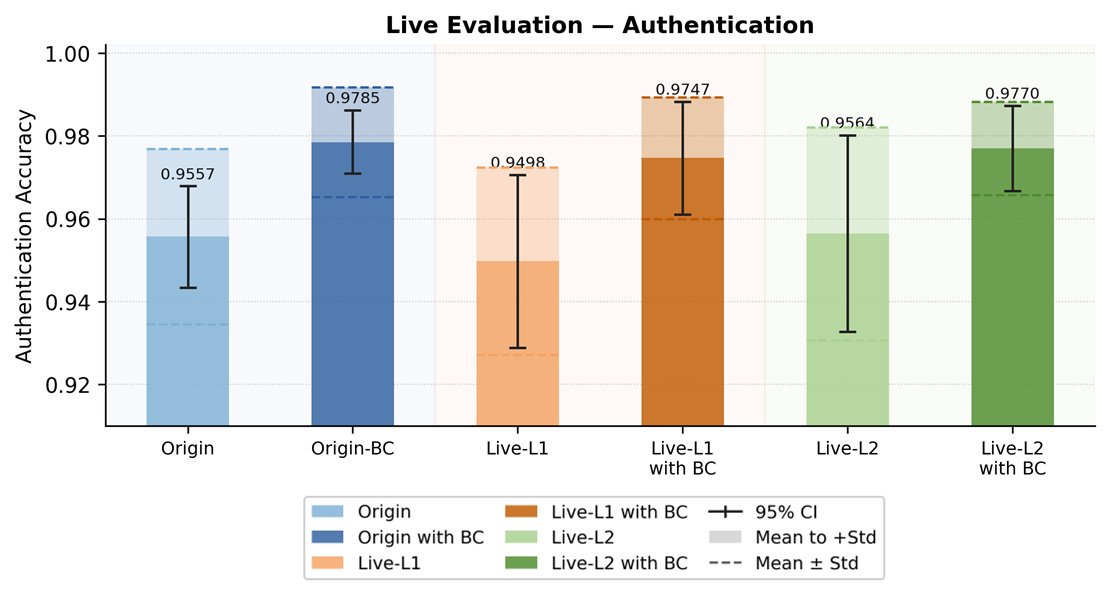}
    \caption{Authentication accuracy under two levels of simplification}
    \label{fig:table18}
\end{figure}

\subsubsection{Liveness Detection and Authentication Accuracy}
We evaluate accuracy on Benchmark Task~1 and Task~3 using 7-fold cross-validation. For liveness detection (Fig.~\ref{fig:table17}), both levels maintain performance close to the original model, with only a slight drop in attack setting~1 due to reduced TDT/ERT precision. For authentication (Fig.~\ref{fig:table18}), no noticeable degradation is observed under either level, with or without BC features, demonstrating strong robustness under simplified processing.

\end{redblock}
\section{Discussion}
In this section, we first summarize the key findings of this work (Section~\ref{sec::findings}) and then highlight the potential applications of the proposed system (Section~\ref{sec::application}). Finally, we discuss its limitations and outline possible directions for future research (Section~\ref{sec:limitation}).

\subsection{Conclusion of Findings}
\label{sec::findings}
In this work, we systematically investigate voice for liveness detection and authentication on smart glasses, leading to the following insights.
\begin{redblock}

\textbf{Physics-driven liveness robustness.}
On-body propagation features (AuthG-Live) generalize to unseen attacks (Section~\ref{sec::benchmark_result_task2}), as replayed audio fails to reproduce spatial and anatomical constraints.

\textbf{Multi-modal necessity for authentication.}
Fusing AC, BC, and SF consistently outperforms unimodal/bimodal baselines across standard and cross-utterance settings (Section~\ref{sec:benchmark_result}), improving robustness to spoofing and intra-speaker variability.

\textbf{Deployment insights under real constraints.}
Performance remains strong under reduced microphone configurations, validating feasibility on commodity devices (Section~\ref{:Evaluation}); meanwhile, we observe that symmetric microphone layouts limit spatial feature capture, highlighting the importance of placement-aware sensor design.

\textbf{Alignment with implicit authentication.}
Passive sensing (no extra user effort), on-body constraints (BC/SF tied to wearing condition), and strong cross-utterance and unseen-attack generalization (Section~\ref{sec:benchmark_result}) together indicate that the learned representations are compatible with implicit and opportunistic authentication pipelines, where authentication can be triggered seamlessly during natural voice interaction.
\end{redblock}

\subsection{Potential Applications}
\label{sec::application}
\subsubsection{AuthG-Live and AuthG-Net}

\red{AuthG-Live and AuthG-Net enable secure voice-based interaction on smart glasses by leveraging multi-acoustic cues for joint liveness detection and authentication. The system can handle diverse real-world voice inputs, including those from the wearer, surrounding users, or when the device is not worn.}
By integrating authentication into passive multi-acoustic sensing, the system naturally supports seamless and implicit verification during normal voice interaction, without requiring additional user effort. In combination with other on-device sensors, such as a camera, this enables intuitive and secure interaction scenarios. For example, a user may simply look at a payment QR code and issue a voice command such as 'Pay the bill,' \red{with authentication performed transparently as part of the interaction.}

\subsubsection{AuthGlass Dataset}

As demonstrated in Section~\ref{sec:benchmark_tasks_design}, the AuthGlass Dataset is designed to facilitate comprehensive evaluation of voice liveness detection and authentication on smart glasses. 
By providing synchronized multi-channel audio inputs, the dataset enables systematic ablation studies across different acoustic modalities and input channel configurations, allowing researchers to investigate the contribution of each modality and channel to system performance. 
Moreover, the dataset includes both genuine samples and attack data collected from multiple users, which supports rigorous evaluation of authentication and liveness detection under diverse adversarial scenarios.

\subsection{Limitation and Future Works}
\label{sec:limitation}
\subsubsection{Dataset Bias and Generalizability}
\label{sec::chap5_limit_dataset}
\red{
Although AuthGlass includes 42 participants and multiple attack settings, it may still exhibit dataset bias due to the relatively homogeneous participant pool at university campus and controlled collection conditions. This may limit generalization to broader populations (e.g., diverse demographics, accents, and speaking styles) and real-world scenarios. In addition, the predefined attack settings may not cover all adversarial strategies in practice. To address this limitation, we open-source both the dataset and the prototype to facilitate future data collection and enrichment across more diverse users, environments, and attack conditions.
}

\subsubsection{More Adversarial Scenarios}
\label{sec::chap5_more_attacks}

\red{
While our dataset covers representative attack settings derived from the defined threat model, it does not exhaust all possible adversarial strategies. More advanced attacks, such as adaptive replay, multi-source injection, or high-fidelity voice synthesis, may better approximate genuine acoustic patterns and are not fully captured in our current evaluation. In addition, variations in wearing conditions may introduce ambiguity in on-body acoustic cues.
To address this limitation, we release the AuthGlass dataset and hardware design to enable future work to explore more diverse and challenging adversarial scenarios and improve system robustness.
}

\subsubsection{In-the-Wild Data and Noise}
Although the AuthGlass dataset is relatively large, it is collected under controlled conditions and does not fully capture real-world acoustic variability. In practical deployment, environmental noise may degrade performance. \red{While fallback mechanisms can mitigate usability issues caused by false rejections, we do not explicitly evaluate or optimize robustness under noisy conditions}, nor do we include noise in data collection. To address this limitation, we open-source our prototype and dataset to support future data collection in real-world environments. Additionally, noise augmentation techniques (e.g., using multi-channel noise datasets~\cite{thiemann2013demand}) can be explored to improve robustness.

\subsubsection{More Configurations of AC Microphones}
Our study validates the effectiveness of incorporating sound-field features for liveness detection and authentication. However, we only explore a limited set of AC microphone configurations and do not fully optimize microphone placement. While our results suggest that spatial distribution plays a critical role, a systematic exploration of layout design remains missing. By releasing our dataset and prototype, we aim to facilitate future studies on microphone placement and configuration strategies, which may further improve performance and provide deeper insights into sound-field sensing on smart glasses.

\subsubsection{Deployment to Commercial Devices}
Despite demonstrating strong performance, several challenges remain for real-world deployment. The computation of sound-field features (e.g., TDT, ERT, EDF) introduces additional processing overhead and requires accurate synchronization across microphones. While our simulated real-time implementation demonstrates that input-independent processing can be achieved with a constant latency of approximately 1.5 seconds, further optimization is still needed for efficient deployment on resource-constrained commercial devices. In addition, integrating bone-conduction sensors into compact and wearable designs remains an open hardware challenge. Nevertheless, ongoing advances in embedded systems and sensor integration are likely to improve the feasibility of deploying such systems in practice.

\subsubsection{Implicit Authentication}
\red{Voice interaction is inherently passive, and our approach further leverages on-body constraints through BC and SF features. Combined with strong generalization in both liveness detection and authentication, as well as low-latency processing in the live setting, these properties align with key requirements of implicit authentication systems. However, our current design and evaluation follow an event-driven pipeline, and both AuthG-Live and AuthG-Net are formulated accordingly. By releasing our dataset and prototype, we aim to enable future work to extend this framework toward continuous and implicit authentication under real-time audio streams.}

\subsubsection{Sound Field Based Interaction on Smart Glasses}
Our findings indicate that sound-field features are closely related to human speech production and propagation, suggesting broader interaction possibilities. Prior work such as Proximic~\cite{qin2021proximic} and NasoVoce~\cite{rekimoto2026nasovoce} uses inter-channel differences to detect mouth-related gestures. Building on this insight, sound-field features on smart glasses may enable gesture recognition (e.g., hand-to-mouth interactions) during speech without requiring active acoustic sensing~\cite{Echotouch}, opening new directions for passive and multimodal interaction.

\section{Conclusion}
This work presents AuthGlass, along with two methods, AuthG-Live and AuthG-Net, for voice liveness detection and authentication on smart glasses. Our results show that combining sound-field and multimodal acoustic features enables robust defense against diverse spoofing and impersonation attacks while maintaining high authentication accuracy. Beyond explicit verification, our results suggest the potential of leveraging voice interaction as an implicit, event-driven authentication signal, enabling seamless and hands-free security without additional user effort.
We further contribute AuthGlass, the first multi-acoustic-modal, multi-channel dataset tailored to smart glasses, providing a unified benchmark for evaluating liveness detection and authentication under realistic threat scenarios. By releasing this dataset and benchmark, we aim to facilitate reproducible research and future exploration of practical voice-based security systems, including continuous or hybrid authentication, cross-device generalization, and resource-aware deployment on commodity smart glasses.

\section{Use of AI Tools}
We use GPT-5 for grammar correction and sentence refinement. We also use GPT-5 and Claude Code to beautify picture. To be specific, sketches in Fig. 1, Fig.~2a, Fig.~3b, Fig.~8b, are generated by GPT-5. Plots Fig. 11 to 17, Fig. 19, 20, 22, 23, 24, 25 are polished by Claude Code.

\begin{acks}
This work is supported by the National Key R\&D Program of China under Grant No. 2024YFB4505500 \& 2024YFB4505503. This work is supported by Ant Group.
\end{acks}

\bibliographystyle{ACM-Reference-Format}
\bibliography{reference}
\appendix
\section{Appendix on Dataset and Methods for Benchmarking}

\subsection{Command Set and Rationale}
\label{sec:passphrase_set}
Table~\ref{tab:passphrase_set} lists the 15 commands used in our experiments. Each phrase contains no more than five words, enabling the evaluation of liveness detection and speaker authentication. They were selected from common daily expressions that could serve as practical voice commands, ensuring the scenarios are realistic for everyday use. In addition, the set was chosen to collectively cover most English vowels and consonants, providing representative phonetic content so that the model’s performance is not biased by the specific phonemes of individual phrases.

\begin{table}[t]
\centering
\caption{Passphrase Set Used in the Experiments}
\label{tab:passphrase_set}
\begin{tabular}{lll}
\hline
(1) Hi siri &
(2) Hi google &
(3) Hi alexa \\
(4) Pay the bill &
(5) Call tom &
(6) Turn on bluetooth \\
(7) Take a photo &
(8) Play some music &
(9) Show me my first message \\
(10) Send an email to john &
(11) Check my voicemail &
(12) What's my appointment \\
(13) Set an alarm &
(14) Thank you &
(15) A hundred dollars \\
\hline
\end{tabular}
\end{table}

\subsection{Additional Participants Demography}
\label{app:demographic}
\begin{figure}
    \centering
    \includegraphics[width=1.0\linewidth]{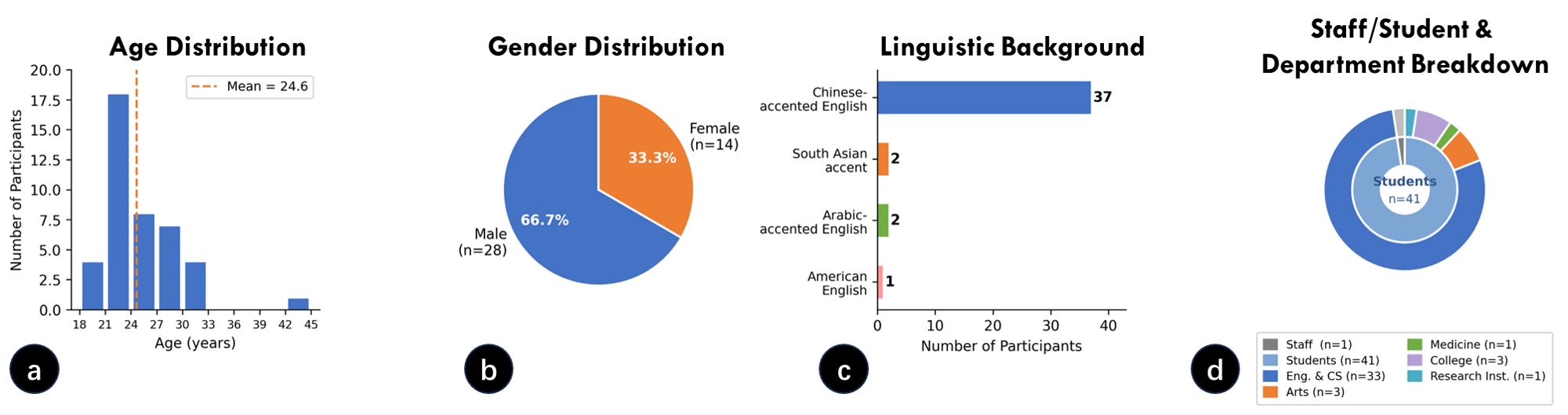}
    \caption{Demographic overview of the 42 participants in the AuthGlass dataset.
    (a)~Age distribution (range: 18–42, mean $24.6 \pm 4.3$ years).
    (b)~Gender distribution: 28 male (66.7\%) and 14 female (33.3\%).
    (c)~Linguistic background: the majority are non-native English speakers
    with Chinese-accented English ($n=37$), alongside speakers with
    South Asian ($n=2$), Arabic-accented ($n=2$), and American English ($n=1$) backgrounds,
    reflecting moderate linguistic diversity within a primarily university-campus population.
    (d)~Affiliation breakdown: 41 students and 1 staff member,
    with students drawn predominantly from Engineering \& Computer Science ($n=33$),
    and smaller representation from Arts ($n=3$), interdisciplinary Colleges ($n=3$),
    Medicine ($n=1$), and a Research Institute ($n=1$).}
    \label{fig:demographic}
\end{figure}

We recruited a total of 42 participants from a university campus community (Fig.~\ref{fig:demographic}). The age of participants ranged from 18 to 42 years, with a mean of $24.6 \pm 4.3$ years. Among them, 28 identified as male (66.7\%) and 14 as female (33.3\%).

The participant pool primarily consisted of students (n=41), with one staff member. Students were drawn from diverse academic backgrounds, including Engineering and Computer Science (n=33), Arts (n=3), interdisciplinary Colleges (n=3), Medicine (n=1), and a Research Institute (n=1), reflecting a typical campus population.

In terms of linguistic background, the majority of participants were non-native English speakers using Chinese-accented English (n=37). Additional accents included South Asian (n=2) and Arabic-accented English (n=2), along with one native speaker of American English (n=1). While the dataset is dominated by a single accent group, it still captures moderate linguistic variability within a realistic university setting.

All participants reported being proficient English users in daily communication. During data collection, they were instructed to speak at a natural conversational pace, maintain clear pronunciation, and avoid exaggerated articulation. This setup ensures that the recorded speech reflects realistic voice-command usage rather than carefully controlled or artificially slowed speech.

Data collection was conducted in a quiet indoor environment. Participants were asked to remain in a natural seated posture while wearing the prototype device. No strict constraints were imposed on individual speaking styles beyond clarity and naturalness, allowing natural inter-speaker variability in speech patterns.

\subsection{Selected Methods for Benchmarking and Our Modifications}
\label{app:method_selection}
\paragraph{Liveness Detection.}
For liveness detection, we selected 7 methods for benchmarking.

\textbf{VOID~\cite{ahmed2020void}}  
VOID extracts four types of features from a single-channel AC signal, including three handcrafted features (low-frequency power, signal power linearity, and high-frequency power) and LPCCs. These features are fed into several traditional machine learning classifiers, among which an SVM with an RBF kernel achieves the best performance. To extend VOID to multi-channel inputs, we reapply the entire feature extraction pipeline to each input channel and concatenate all channel-wise features as the input to the SVM classifier.

\textbf{He et al.~\cite{he2024fast}}  
Similar to VOID, this method extracts five types of features from a single-channel AC signal, consisting of four handcrafted features and LPCCs. The extracted features are input to a self-designed SE-ResNet, a residual convolutional neural network (CNN). To support multi-channel inputs, we perform feature extraction independently on each channel and stack the resulting features across channels to form a multi-channel input. We modified the input layer of the SE-ResNet for the multi-channel input.

\textbf{Boneauth \cite{li2024boneauth}} Boneauth computes MFCCs from both BC and AC signals and performs liveness detection by measuring the cosine similarity between the two modalities. An input is accepted only if the similarity exceeds a predefined threshold. The threshold is determined by minimizing the Equal Error Rate (EER). Since the original method operates on a single BC–AC pair, we extend it to the multi-channel setting by pairing the BC channel with each AC channel individually and applying majority voting to determine the final liveness decision.

\textbf{Eve Said Yes~\cite{huang2025eve}}  
Eve Said Yes uses pooled STFT energy spectrograms as input features and defines a temporal similarity metric between AC and BC signals for liveness detection. An utterance is accepted if the similarity exceeds a threshold, determined by minimizing EER. As this method also relies on a single AC–BC channel pair, we adopt the same multi-channel voting strategy as used for Boneauth.

\textbf{Li et al.~\cite{li2021robust}}  
This method is inherently designed for multi-channel voice liveness detection. It leverages both energy and phase spectra derived from STFT as input features and employs a CNN for classification. Since the model naturally accommodates multi-channel inputs, we directly feed features from all selected channels only with the modification on the CNN input layer.

\textbf{Cafield~\cite{yan2019catcher}}  
Cafield proposes Fieldprint, a feature computed as the ratio of the energy–frequency relationship vectors between two channels. A GMM is trained to model genuine samples and distinguish them from attack samples. We extend this method by computing Fieldprints between each channel and the reference AC channel (Channel~7), and concatenating all resulting features for GMM-based classification

\textbf{Voshield~\cite{yang2023voshield}}  
Voshield utilizes sound field dynamics (SFD), defined as the ratio of STFT spectra between two channels, and employs a residual CNN for binary classification. Similar to Cafield, we compute SFD features between each channel and the reference AC channel (Channel~7), and stack all resulting features as the input to the CNN, with its' input layer modified for all the input channels.

\paragraph{Authentication.}
For authentication, we selected 4 methods for benchmarking.

\textbf{Park et al.~\cite{park2025toward}}  
This method originally operates on a single AC channel and uses MFCCs as input features. A self-designed LSTM network is employed as the authentication model. During enrollment, embeddings extracted from an intermediate layer are stored as user templates, and cosine similarity is used to compare test embeddings against the enrolled template. Authentication is accepted if the similarity exceeds a predefined threshold. Following this procedure, we compute MFCC features for all selected channels, stack them together, and modify the model input layer accordingly to support multi-channel inputs.

\textbf{Eve Said Yes~\cite{huang2025eve}}  
Eve Said Yes uses CQT features extracted from a single BC channel as input to a residual CNN trained with adversarial learning to produce speaker embeddings. During enrollment, embeddings are extracted from an intermediate fully connected layer and compared with stored templates using cosine similarity. Similar to Park et al., we extend this method by stacking features extracted from all channels and adapting the model input layer to accommodate the expanded multi-feature input.

\textbf{Cafield~\cite{yan2019catcher}}  
As discussed previously, Cafield employs a GMM-based framework for both liveness detection and authentication. We apply the same multi-channel feature construction and model modification strategy as used in the liveness detection setting.

\textbf{Voshield~\cite{yang2023voshield}}  
Voshield does not originally support authentication, as it is designed for binary liveness classification. However, since it is based on a CNN architecture, it can be readily adapted for authentication by replacing the binary output layer with a fully connected layer followed by a softmax output. We extract embeddings from the fully connected layer immediately following the residual block and perform authentication using cosine similarity, following the same protocol as Park et al. and Eve Said Yes.

\section{Detailed Implementation of Feature Extraction and AuthG-Net}
\subsection{Feature Extraction Implementation}
\label{app:feature_construction}

This appendix provides a detailed description of the feature extraction procedures used for both air-conduction (AC) and bone-conduction (BC) signals, as well as the derivation of sound field (SF) features.

\subsubsection{AC and BC Channels}

Log-Mel spectrograms were computed for both AC and BC channels. For AC signals, raw waveforms were directly transformed into Mel spectrograms. In contrast, BC signals were first processed using a Butterworth high-pass filter with a cutoff frequency of 100~Hz to suppress motion-induced low-frequency artifacts and ambient noise. 

All signals were subsequently resampled to 96~kHz via linear interpolation to ensure a consistent temporal resolution. Mel spectrograms were then generated using 128 Mel filter banks with an upper frequency limit of 8~kHz. The resulting representations capture perceptually relevant spectral characteristics while maintaining a uniform feature dimensionality across modalities.

\subsubsection{Sound Field Feature Construction}
To characterize spatial acoustic properties, three complementary features were extracted from the AC SF channels: Energy Ratio over Time (ERT), Time Delay over Time (TDT), and Energy Distribution over Frequency (EDF). These features follow the design principles described in Section~\ref{sec::chap3_validation_acoustic}.

\paragraph{Energy Ratio over Time (ERT)}
ERT was computed by first applying the short-time Fourier transform (STFT) to each voiced segment. Spectral energy within the 100--8{,}000~Hz band was then summed for each AC channel (see Section~\ref{sec::chap3_multi-modal_acoustic_data_collection}). To reduce variability caused by absolute sound intensity, channel-wise energy was normalized with respect to a reference channel (Channel~7). The resulting ratios were linearly scaled and further normalized using a hyperbolic tangent ($\tanh$) function to improve numerical stability and bound the feature range.

\paragraph{Time Delay over Time (TDT)}
TDT was derived by estimating relative time delays between each AC channel and the reference channel (Channel~7) using cross-correlation. For each channel, the time lag corresponding to the maximum correlation peak was selected as the delay estimate. A confidence score was computed by normalizing the peak correlation with respect to signal energies and applying a logit-like transformation. The final delay value was obtained by weighting the estimated lag with this confidence score, which suppresses unreliable estimates caused by noise or absent signals while preserving robust temporal alignment cues. As with ERT, the resulting values were linearly scaled and normalized using a $\tanh$ function.

\paragraph{Energy Distribution over Frequency (EDF)}
EDF was computed by applying STFT to each voiced segment across all channels. For each AC channel, the resulting spectrogram was averaged along the temporal dimension to obtain a frequency-wise energy profile. Channel~7 was again used as the reference, and all other channels were normalized by the maximum STFT energy of the reference channel. This normalization preserves relative spectral distributions across frequency bands while mitigating variations in overall signal amplitude.
A summary of all feature-related parameters is provided in Table~\ref{tab:feature_params}.

\begin{table}[t]
\centering
\caption{Feature Extraction Parameters}
\label{tab:feature_params}
\begin{tabular}{lll}
% \begin{tabular*}{\textwidth}{@{\extracolsep{\fill}} l l l}

\hline
\multicolumn{3}{c}{\textbf{Parameters for Mel Spectrogram}} \\[2pt]
Sampling rate: 96,000 Hz &
Maximum frequency: 8,000 Hz&
  \\
Hop length: 1024 &
$N_{FFT}$: 2048 &
$N_{MELS}$: 128 
\\
\hline
\multicolumn{3}{c}{\textbf{Parameters for ERT and TDT}} \\[2pt]
Sample rate: 96,000 Hz &
Frequency range: 100--8,000 Hz 
\\
Step length: 0.005 s &
Scaling factor: 2000 
 \\
 \hline
\multicolumn{3}{c}{\textbf{Parameters for EDF}} \\[2pt]
Sample rate: 96,000 Hz &
Maximum frequency: 8,000 Hz&
  \\
Hop length: 1024 &
$N_{FFT}$: 1024 
\\

\hline
\end{tabular}
\end{table}
\subsection{AuthG-Net Implementation}
\label{app:authG-Net_implementation}
AuthG-Net is a lightweight multimodal neural architecture for user authentication (see Fig.~\ref{fig:AuthG_model}). It jointly exploits AC, BC, and SF modalities to learn discriminative and passphrase-independent user representations. The model follows a three-stage pipeline: modality-specific encoding, feature fusion, and embedding learning with adversarial supervision.

Each modality is first processed by an independent encoder. For the AC and BC inputs, mel-log spectrogram features (Section~\ref{sec::our_method_feature_extraction}) are used. All samples are zero-padded to a fixed size of $(215, 128)$ to ensure consistent input dimensions across the dataset. Each encoder consists of a single $3\times3$ 2D convolution layer (stride 1, 32 output channels) followed by a ReLU activation. Since AC and BC share identical feature extraction settings, the same encoder architecture is adopted for both.

For the SF modality, three complementary features: TDT, ERT, and EDF are jointly utilized. TDT and ERT are naturally aligned with the temporal resolution of AC/BC features and are padded to $(215, 13)$. To mitigate instability in TDT when the signal energy is low, a masking strategy is applied based on the AC signal energy, zeroing out frames below a predefined threshold. EDF, originally of size $(86, 13)$, is temporally rescaled via linear interpolation to $(215, 13)$. The three SF features are concatenated along the temporal dimension and passed through a $3\times3$ 2D convolution layer (stride 1, 32 channels), followed by a linear projection that expands the feature width to 128. ReLU activations are applied throughout.

The encoded features from AC, BC, and SF branches are concatenated along the channel dimension, producing a fused representation with 96 channels. This fused tensor is processed by a modified ResNet-18 backbone, where the first convolution layer is adapted to accept the increased number of input channels. The backbone retains all four residual blocks and outputs a 512-dimensional embedding after global average pooling and flattening.

To enforce passphrase-invariant representation learning, a domain adversarial learning (DAL) framework~\cite{tzeng2017adversarial, ganin2015unsupervised} is employed. Two classifiers are attached to the embedding: a speaker classifier and a passphrase (content) classifier, each implemented as a three-layer fully connected network with a softmax output. The speaker classifier is optimized directly, while the passphrase classifier is connected via a Gradient Reversal Layer (GRL), which inverts gradients during backpropagation to suppress content-related information. The overall training objective is given by:
\[
L(w_f, w_s, w_p) = L_s(w_f, w_s) - \lambda \cdot L_p(w_f, w_p),
\]
where $L_s$ and $L_p$ denote cross-entropy losses for speaker and passphrase classification, respectively, and $\lambda=1$. The parameters $w_f$, $w_s$, and $w_p$ correspond to the feature extractor, speaker classifier, and passphrase classifier.

During training, each sample is annotated with both user identity and utterance labels, and the model is optimized using Adam. After convergence, the 512-dimensional embedding (highlighted in Fig.~\ref{fig:AuthG_model}) is used as the user representation. For enrollment, embeddings are extracted and stored as templates. During authentication, cosine similarity is computed between the stored template and the query embedding; access is granted if the similarity exceeds a predefined threshold.

\section{Results for Experiments}

\subsection{Performance of the Original Implementation of Liveness Detection Methods}
\label{sec:performance_of_the_original_liveness}

\begin{figure}
    \centering
    \includegraphics[width=0.8\linewidth]{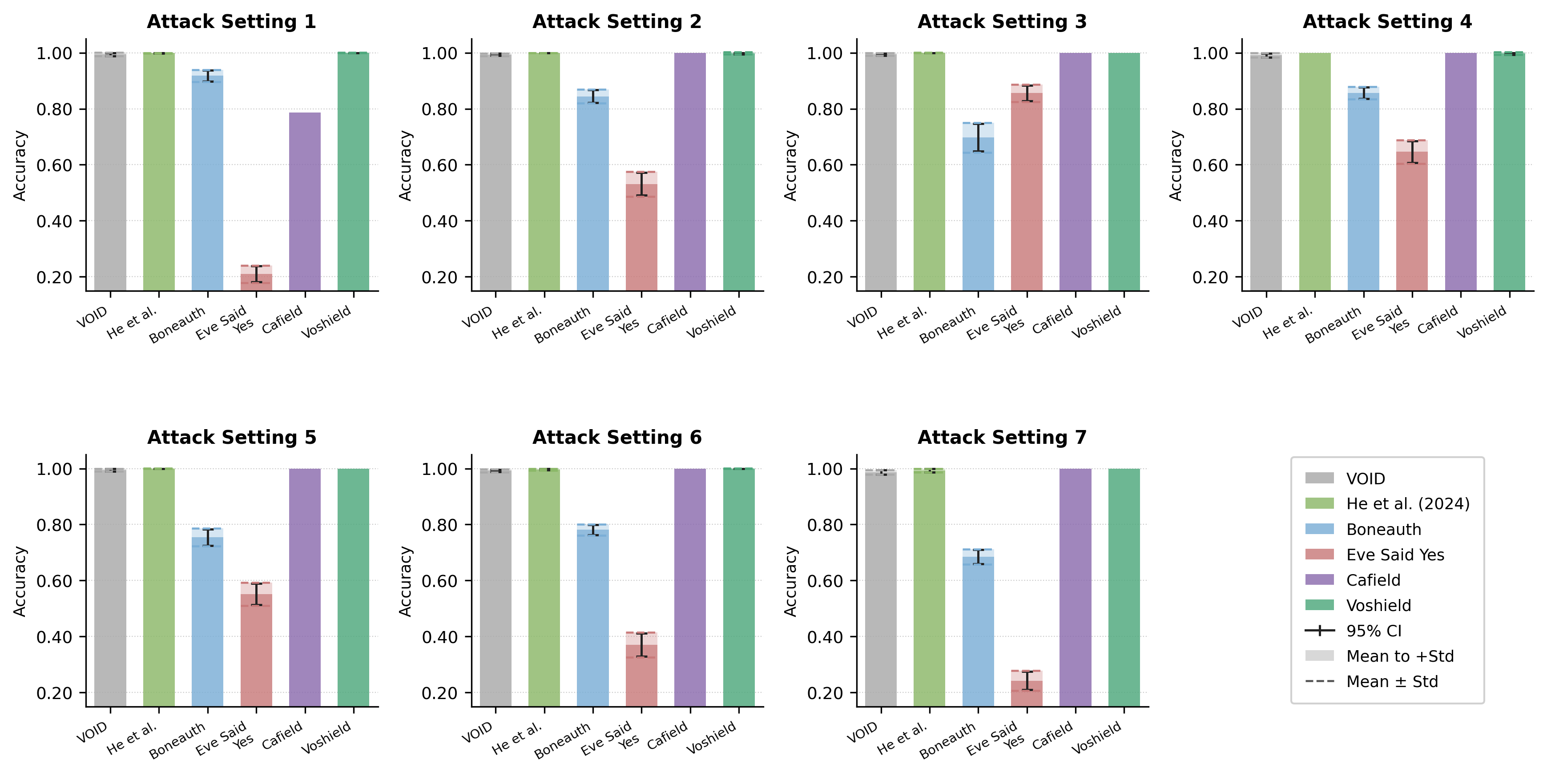}
    \caption{Liveness detection accuracy for original baseline implementation}
    \label{fig:appendix_liveness}
\end{figure}

In Section 4.5, we evaluated the accuracy of previous methods on multi-channel microphone data in Benchmark Task 1. In some of the original papers, only a small number of microphones were used. Based on the descriptions in the original works, we also implemented the original versions of the selected methods. The liveness detection accuracy results are presented in Fig.~\ref{fig:appendix_liveness}.

% \begin{table}[htbp]
% \centering
% \caption{Comparison on Liveness Detection Accuracy Using the Original Versions of the Selected Methods.}
% \label{tab:attack_comparison}

% \small
% \setlength{\tabcolsep}{5pt}
% \renewcommand{\arraystretch}{1.15}

% \begin{threeparttable}
% \begin{tabular}{lccccccc}
% \toprule
% \textbf{Attacked by} 
% & \textbf{Attack 1} 
% & \textbf{Attack 2} 
% & \textbf{Attack 3} 
% & \textbf{Attack 4} 
% & \textbf{Attack 5} 
% & \textbf{Attack 6} 
% & \textbf{Attack 7} \\
% \midrule
% VOID               & 99.43\% & 99.14\% & 99.21\% & 98.81\% & 98.69\% & 98.56\% & 96.60\% \\
% Fast and Lightweight
%                    & 99.92\% & 99.95\% & 99.97\% & 100.00\% & 99.99\% & 99.64\% & 99.29\% \\
% Boneauth           & 91.80\% & 84.42\% & 69.70\% & 85.62\% & 75.41\% & 78.11\% & 68.44\% \\
% Eve Said Yes       & 20.95\% & 53.15\% & 85.58\% & 64.60\% & 55.13\% & 37.01\% & 24.21\% \\
% Cafield            & 78.69\% & 100.00\% & 100.00\% & 100.00\% & 99.96\% & 100.00\% & 100.00\% \\
% Voshield           & 100.00\% & 100.00\% & 100.00\% & 100.00\% & 100.00\% & 100.00\% & 100.00\% \\
% \bottomrule
% \end{tabular}
% \end{threeparttable}
% \end{table}

\end{document}